\documentclass[12pt]{article}

\usepackage{graphicx}
\usepackage{natbib}
\usepackage{multirow}%
\usepackage{amsmath,amssymb,amsfonts}%
\usepackage{amsthm}%
\usepackage{mathrsfs}%
\usepackage[title]{appendix}%
\usepackage{textcomp}%
\usepackage{manyfoot}%
\usepackage{booktabs}%
\usepackage{listings}%
\usepackage{subcaption}

\newcommand{\vect}[1]{\boldsymbol{#1}}

\begin{document}

\title{Estimation of a multivariate von Mises distribution for contaminated torus data}
\author{
Giulia Bertagnolli \\
Faculty of Economics and Management, \\
Free University of Bozen-Bolzano, Italy \\
and \\
Luca Greco \\
University G. Fortunato, Benevento, Italy \\
and \\
Claudio Agostinelli \\
Department of Mathematics, University of Trento, Italy}

\maketitle

\begin{abstract}
The occurrence of atypical circular observations on the torus can badly affect parameter estimation of the multivariate von Mises distribution.
This paper addresses the problem of robust fitting of the multivariate von Mises model using the weighted likelihood methodology.
The key ingredients are non-parametric density estimation for multivariate circular data and the definition of appropriate weighted estimating equations.
Some theoretical properties are discussed. The  finite sample behavior of the proposed weighted likelihood estimator has been investigated by Monte Carlo numerical studies and empirical applications.
\end{abstract}


\section{Introduction}
\label{sec:1}
In many contexts, the interest consists in studying the joint distribution of $p>1$ circular observations recorded on a periodic scale, such as wind directions taken in different locations  \citep{lund1999cluster, agostinelli2007robust}, angles of human or robotic arm movements \citep{navarro2017multivariate}, consecutive torsion angles in protein structures \citep{mardia2007protein, mardia2012mixtures, Eltzner2018, greco2023finite}.  Among the relevant applications we also find the analysis of animal movements \citep{ranalli2020model}, 
handwriting recognition \citep{bahlmann2006directional}, people orientation \citep{baltieri2012people}, cognitive and experimental psychology \citep{warren2017wormholes}, human motor resonance \citep{cremers2018one}, neuronal activity \citep{rutishauser2010human}, applications in astronomy, meteorology, geology \citep{lark2014modelling}, medicine, oceanography \citep{jona2012spatial}, physics, image and audio modelling \citep{wadhwa2013phase}.

In a multivariate setting, a circular measurement is represented by a point on the unit circle in each dimension. Then, multivariate circular data can be thought of as points on a torus $\mathbb{T}^p = [0, 2\pi)^p$ (or any other interval of length $2\pi$), whose surface is obtained by revolving the unit circle in a $p$-dimensional manifold, that is considering the Cartesian product of $p$ unit circles.
The modeling of circular data has been tackled through different suitable distributions on the torus, such as the von Mises distribution \citep{mardia1972statistics, mardia2012statistics, mardia2012mixtures}. The Wrapped Normal and the Projected Gaussian distribution are popular alternatives \citep[see][for a review]{MardiaJupp2000, pewsey2013circular}. The key element in understanding circular data lies in their periodicity, as the angles $\vect{\theta} + 2\pi \vect{j}$, with $\vect{j}\in\mathbb{Z}^p$, are the same data point on the torus. Therefore, estimation and inference should reflect this aspect.

The main objective of this work is to develop a reliable estimation process of the parameters of a multivariate von Mises distribution when outlying circular data occur. Outliers are unexpected anomalous values that exhibit a pattern different from the rest of the data \citep[see][for a gentle introduction to robustness]{farcomeni2016robust}, as in the case of data orientated towards certain rare directions in some dimensions or because of the presence of hidden subgroups and structures in the data \citep{agostinelli2024weighted}.
Circular outliers can very badly affect likelihood based estimation leading to unreliable inferences, and robust estimation is supposed to mitigate their adverse effects. Specifically, here we recommend using the weighted likelihood methodology.
The first results about robust fitting of the von Mises model using the weighted likelihood can be found in \cite{agostinelli2007robust} in the univariate setting.
In the multivariate framework \cite{saraceno2021robust, greco2021robust, agostinelli2024weighted} address the problem of robust estimation for circular data under the Wrapped Normal model assumption. Results for the multivariate von Mises distribution are, however, still missing, despite the centrality of this distribution in many fields of application.

The remainder of this paper is organized as follows: some background on the multivariate von Mises distribution is given in Section \ref{sec:2}; we review some fundamental results about weighted likelihood estimation in Section \ref{sec:3}; robust fitting is described in Section \ref{sec:4};
numerical studies are presented in Section \ref{sec:5}; empirical applications are given in Section \ref{sec:6}; concluding remarks end the paper in Section \ref{sec:7}.

\section{The multivariate von Mises distribution}
\label{sec:2}
Let $\vect{\theta}=(\theta_1, \theta_2, \ldots, \theta_p)$ be a $p$-variate circular random variable. 
The von Mises sine distribution~\citep{Mardia2008} has density function
\begin{equation}
	\label{eq:dvmsin}
	m(\vect{\theta}; \vect{\mu}, \vect{\kappa},\Lambda) = C^{-1}( \vect{\kappa},\Lambda)\exp\left[\vect{\kappa}^\top \cos(\vect{\theta}-\vect{\mu})+\frac{1}{2}\sin(\vect{\theta}-\vect{\mu})^\top \Lambda \sin(\vect{\theta}-\vect{\mu})\right]
\end{equation}
with $\vect{\theta} \in [0, 2\pi)^p$, location vector $\vect{\mu} = (\mu_1, \ldots,\mu_p)^\top \in [0, 2\pi)^p$, concentration vector $\vect{\kappa} = (k_1, \ldots, k_p)^\top$ with $k_j>0$, $\Lambda$ is a $p\times p$ symmetric matrix with null diagonal elements $\Lambda_{jj}= 0$ and off-diagonal entries $\Lambda_{ij}=\lambda_{ij}=\lambda_{ji}$, $i,j=1,\ldots,p$.
We note that for $p=1$, \eqref{eq:dvmsin} gives the univariate von Mises density. 
For $p=2$ the density in \eqref{eq:dvmsin} corresponds to the bivariate model of \cite{singh2002probabilistic}:
in this situation, the $\Lambda$ matrix is characterized by the only non-null entry $\lambda$.

\begin{figure}[ht]
	\centering
	\includegraphics[width=0.6\textwidth]{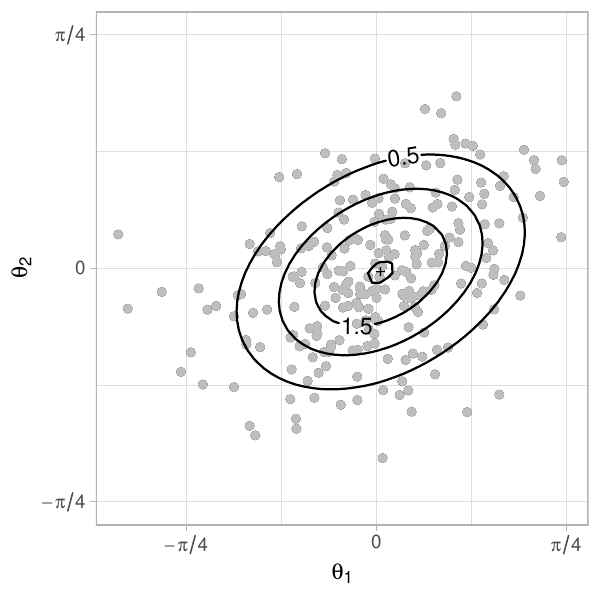}
	\caption{Sample of size $n=250$ from a bivariate von Mises distribution with $\boldsymbol{\kappa}=(10,20)$ and $\lambda=5$. Fitted density contours obtained using approximate maximum likelihood estimates superimposed.}
	\label{fig:1}
\end{figure}

\begin{figure}[ht]
	\centering
	\includegraphics[width=0.45\textwidth]{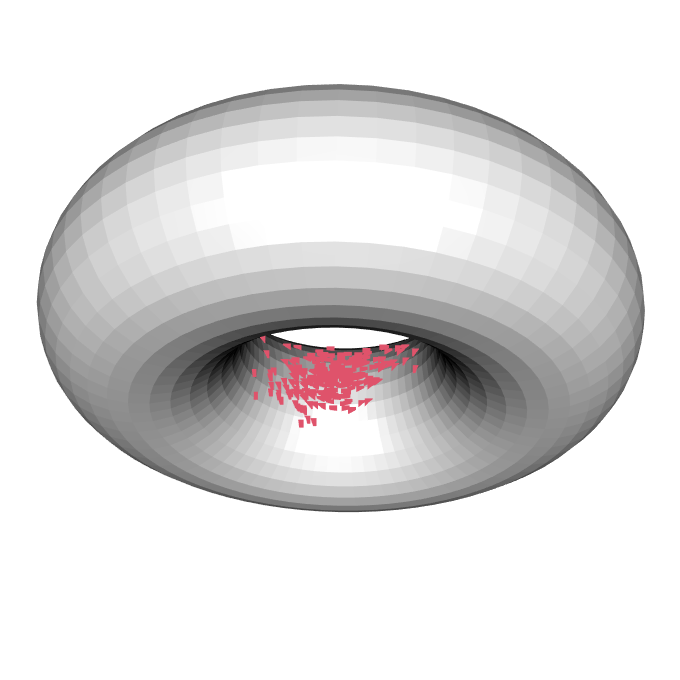}
	\includegraphics[width=0.45\textwidth]{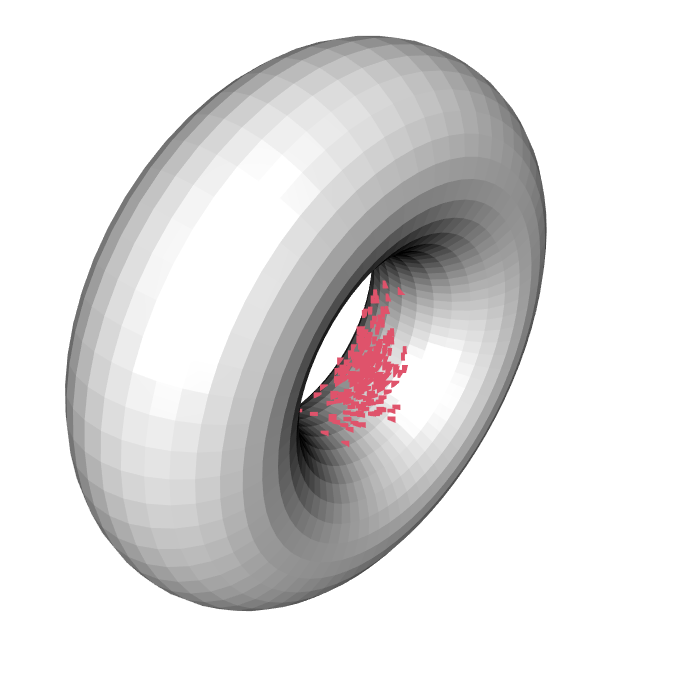}
	\caption{Genuine data. Sample of size $n=250$ from a bivariate von Mises distribution with $\vect{\kappa}=(10,20)$ and $\lambda=5$. Data are displayed as point on $\mathbb{T}^2$ from different perspectives.}
	\label{fig: torus-genuine}
\end{figure}

The normalizing constant $C^{-1}( \vect{\kappa},\Lambda)$ is unknown in any explicit form for $p>2$, but can be approximated as
\begin{equation}\label{eq:capprox}
	C( \vect{\kappa},\Lambda)\approx (2\pi)^{\frac{p}{2}}|\Sigma|^{\frac{1}{2}}\exp\left[\sum_{j=1}^p k_j \right]
\end{equation}
for large values of the concentration parameters in $\vect{\kappa}$, given that the matrix $\Sigma$, where $(\Sigma^{-1})_{ij}=-\lambda_{ij}$ and $(\Sigma^{-1})_{jj}=k_{j}$ for $i,j=1, \dots, p$, is positive definite \citep{mardia2008multivariate, mardia2012mixtures}.
For $p=2$, $\Sigma$ is positive definite whenever $\lambda^2 < k_1k_2$~\citep{mardia2012statistics}.
Positive definiteness of $\Sigma$ also guarantees unimodality. 
Henceforth, we assume that $\vect{\kappa}$ and $\Lambda$ are chosen so that $\Sigma$ is positive definite.
For sufficiently large concentration parameters, and hence small fluctuations in the random angles, the density function in \eqref{eq:dvmsin} can then be approximated by the so-called Concentrated Multivariate Sine (CMS) density~\citep{mardia2012mixtures}, which is obtained replacing the normalizing constant with its approximation \eqref{eq:capprox} leading to \begin{equation}
		m^c(\vect{\theta}; \vect{\mu}, \vect{\kappa},\Lambda) = (2\pi)^{-\frac{p}{2}}|\Sigma|^{-\frac{1}{2}}\exp\left[-\vect{\kappa}^\top (1-c(\vect{\theta};\vect{\mu}))+\frac{s(\vect{\theta};\vect{\mu})^\top \Lambda s(\vect{\theta};\vect{\mu})}{2}\right] 
		\label{eq:dcms}
\end{equation}
where $s(\vect{\theta};\vect{\mu})= \sin(\vect{\theta}-\vect{\mu})$ and $c(\vect{\theta};\vect{\mu}) = \cos(\vect{\theta}-\vect{\mu})$.
Inference about $\vect{\mu}, \vect{\kappa}$ and $\Lambda$ can be efficiently pursued and approximated through maximum likelihood based on \eqref{eq:dcms}.
Given sample data $\vect{\theta}^{(i)}, i=1, \ldots, n$, (approximate) maximum likelihood estimates (MLE) are obtained by 
\begin{align}
	\label{esteq}
	\hat{\mu}_j & = \mathrm{atan2} (\bar{U}_j, \bar{V}_j) \ , ~j=1, \dots, p \ , \nonumber \\
	\hat \Sigma_{jj} & = \frac{2}{n}\sum_{i=1}^n \left(1-\cos(\theta^{(i)}_j-\hat{\mu}_j)\right) \ , ~j=1, \dots, p \ , \\
	\hat \Sigma_{jl} & = \frac{1}{n}\sum_{i=1}^n \sin(\theta^{(i)}_j-\hat{\mu}_j)\sin(\theta^{(i)}_l-\hat{\mu}_l) \ , ~l \neq j,~j=1, \dots, p \ , \nonumber  
\end{align}
where $\bar{U}_j, \bar{V}_j$ are the sample means of $\cos(\theta_j)$ and $\sin(\theta_j)$ respectively
\citep{mardia2012mixtures}.
Figure \ref{fig:1} shows a sample of size $n=250$ from a  bivariate von Mises distribution, with null mean marginal directions, $\vect{\kappa}=(10,20)$ and $\lambda=5$. Fitted density contours lines are superimposed on $[-\pi, \pi)^2$, that have been obtained using (\ref{esteq}). Data are displayed as points on $\mathbb{T}^2$ from different perspectives in Figure \ref{fig: torus-genuine}.

\section{Background on weighted likelihood estimation}
\label{sec:3}

Consider a random sample $\vect{y}^{(1)}, \cdots, \vect{y}^{(n)}$ of size $n$ drawn from a random variable $\vect{Y}$ 
with distribution function $F(\vect{y})$ and probability density function $f(\vect{y})$. 
Let $\mathcal{M} =  \{ M(\vect{y}; \vect{\tau}), \vect{\tau} \in T \subseteq \mathbb{R}^d, d \geq 1, \vect{y} \in \mathcal{Y} \}$ denote the assumed parametric model, with corresponding density $m(\vect{y};\vect{\tau})$, and $\hat F_n$ be the empirical distribution function. Assume that the support of $M(\vect{y}; \vect{\tau})$ is the same as $F(\vect{y})$ and independent of $\vect{\tau}$. A measure of the agreement between the {\it true} and assumed model is provided by the Pearson residual function
\begin{equation}
	\label{pearson}
	\delta(\vect{y}) = \delta(\vect{y}; \vect{\tau}, F) = \frac{f(\vect{y})}{m(\vect{y}; \vect{\tau})} - 1 \
\end{equation}
with $\delta(\vect{y})\in [-1,+\infty)$ \citep{lindsay1994, markatou1998weighted}.
Large values of the Pearson residual function correspond to regions of the support $\mathcal{Y}$ where the model fits the data poorly, meaning that $\vect{y}$ is unlikely to occur under the assumed model.
The finite sample counterpart of (\ref{pearson}) can be obtained as
\begin{equation}
	\label{eq:residualfs}
	\delta_n(\vect{y}) = \delta(\vect{y}; \vect{\tau}, \hat F_n) = \frac{\hat f_n(\vect{y})}{m(\vect{y}; \vect{\tau})} - 1 \ ,
\end{equation}
where $\hat f_n(\vect{y})$ is a consistent estimate of the true density $f(\vect{y})$.
In continuous families of distributions, $\hat{f}_n(\vect{y})$ is a non-parametric density estimate based on the kernel function $k(\vect{y};\vect{t},H)$, that is
\begin{equation}
	\label{eq:parametric-density}
	\hat f_n(\vect{y})=\int_\mathcal{Y}k(\vect{y};\vect{t},H) d\hat F_n(\vect{t}) \ ,
\end{equation}
where $H$ is a $p\times p$ bandwidth matrix, often chosen as $h I_p$ where $I_p$ is the identity matrix of dimension $p$ and $h > 0$ plays the role of smoothing parameter.
Moreover, in the continuous case, the model density in \eqref{eq:residualfs} can be replaced by a smoothed model density, obtained by using the same kernel involved in non-parametric density estimation, that is
\begin{equation*}
	\hat m(\vect{y}; \vect{\tau})=\int_\mathcal{Y}k(\vect{y};\vect{t},H)m(\vect{t};\vect{\tau}) \ d\vect{t} \
\end{equation*}
leading to
\begin{equation}
	\label{eq:residualfs2}
	\hat{\delta}_n(\vect{y}) = \hat{\delta}(\vect{y}; \vect{\theta}, \hat F_n) = \frac{\hat f_n(\vect{y})}{\hat m(\vect{y}; \vect{\tau})} - 1 \ .
\end{equation}
Then, a weight in the interval $[0,1]$ is attached to each data point, that is computed based on
\begin{equation}
	\label{weight}
	w(\delta_n(\vect{y})) = \min\left\{ 1, \frac{\left[A(\delta_n(\vect{y})) + 1\right]^+}{\delta_n(\vect{y}) + 1} \right\} \ ,
\end{equation}
where $[\cdot]^+$ denotes the positive part and $A(\delta)$ is a Residual Adjustment Function (RAF, \cite{lindsay1994, basu1994minimum, park+basu+2003}), whose
special role is related to the connections between weighted likelihood and minimum disparity estimation \citep{markatou1998weighted, kuchibhotla2017}.
The weights $w(\delta_n(\vect{y}))$ are meant to be small for those data points that are in disagreement with the assumed model.
In practice, the RAF acts by bounding the effect of those points leading to large Pearson residuals, leading to downweighting them. The function $A(\cdot)$ is assumed to be increasing and twice differentiable in $[-1, +\infty)$, with $A(0) = 0$ and $A^\prime(0) = 1$. The weight function (\ref{weight}) can involve a RAF based on the Symmetric Chi-squared divergence (SCHI), the family of Power divergences or the Generalized Kullback--Leibler divergence (GKL)~\citep{lindsay1994, basu1994minimum, park+basu+2003, park+basu+lindsay+2002}. 
Depending on the RAF, the weights decline smoothly to zero as $\delta_n(\vect{y})\rightarrow\infty$, that is when the observation is an outlier but also as $\delta\rightarrow -1$: in the latter case it is common to refer to inliers.
Since inliers represent a minor issue for data in $p$--dimensional torus, we consider a modified RAF, equal to unity in the interval $[-1,0]$  \citep[see][for details]{saraceno2021robust}.

According to the chosen RAF, the weighted likelihood estimate (WLE) $\tilde{\vect{\tau}}$ is the solution to the set of Weighted Likelihood Estimating Equations (WLEE), defined as
\begin{equation}
	\label{eq:wlee}
	\sum_{i=1}^n w\left(\delta_n\left(\vect{y}^{(i)}\right); \vect{\tau}, \hat{F}_n\right) u\left(\vect{y}^{(i)};  \vect{\tau}\right) = \sum_{i=1}^n w_i u\left(\vect{y}^{(i)};  \vect{\tau}\right) = 0 \ ,
\end{equation}
where $u\left(\vect{y}^{(i)};\vect{\tau}\right)$ is the individual contribution to the score function. 
In general, finding the solution of \eqref{eq:wlee} requires an iterative weighting algorithm. 
Potential outliers can be detected through the inspection of the weights $w_i$. Moreover, the empirical downweighting level $edl=1-\bar w$, where $\bar w$ is the mean of the weights, gives a rough measure of the amount of contamination in the sample at hand.

The corresponding weighted likelihood estimator,  is consistent, asymptotically normal, and fully efficient at the assumed model, under some general regularity conditions pertaining to the model, the kernel and the weight function~\citep{markatou1998weighted, agostinelli2001test, agostinelli2019weighted}. Its robustness properties have been established in~\citep{lindsay1994} in connection with minimum disparity problems.

In finite samples, the robustness/efficiency trade-off of WLE can be tuned by varying the smoothing parameter $h$ controlling \eqref{eq:parametric-density}. Large values of $h$ lead to Pearson residuals all close to zero and weights all close to one and, hence, large efficiency, since $\hat f_n(\vect{y})$ become stochastically close to the smoothed model. On the other hand, small values of $h$ make $\hat f_n(\vect{y})$ more sensitive to the occurrence of outliers and the Pearson residuals become large for those data points that are in disagreement with the model. On the contrary, the shape of the kernel function has a very limited effect.

\section{Robust estimation}
\label{sec:4}
\begin{figure}[!h]
	\centering
	\includegraphics[width=0.4\textwidth]{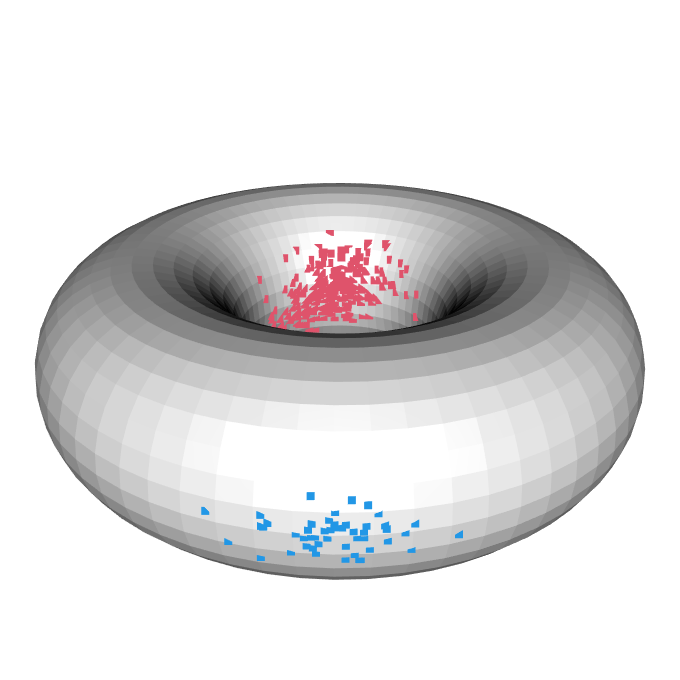} \quad 
	\includegraphics[width=0.4\textwidth]{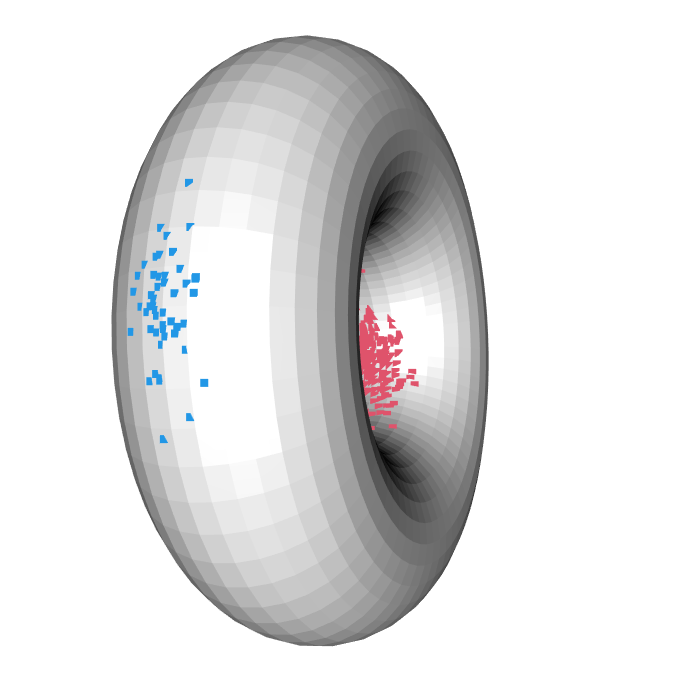} \\
	\includegraphics[width=0.4\textwidth]{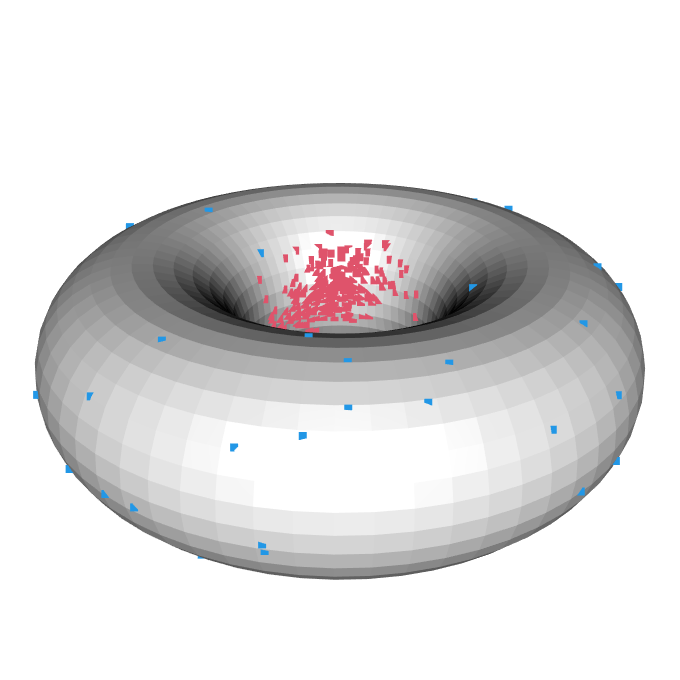} \quad 
	\includegraphics[width=0.4\textwidth]{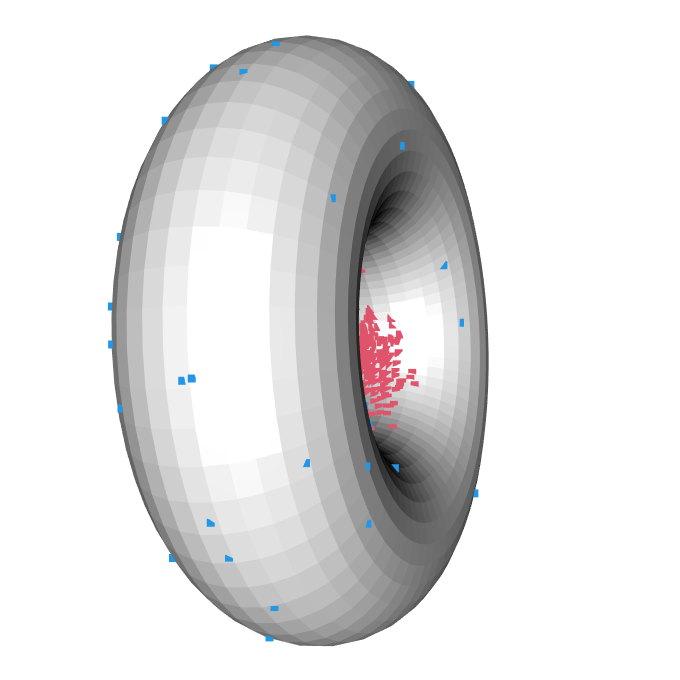} 
	\caption{Contaminated data. Two sets of 50 outliers (blue points) are added to the $n = 250$ genuine data points (see Fig.~\ref{fig: torus-genuine}). The top row shows the case where the outliers are clustered, while the bottom row shows the case where they are scattered. The data are displayed as points on $\mathbb{T}^2$ from different viewing angles.}
	\label{fig: torus-contam}
\end{figure}

\begin{figure}[!ht]
	\centering
	\includegraphics[width=0.45\textwidth]{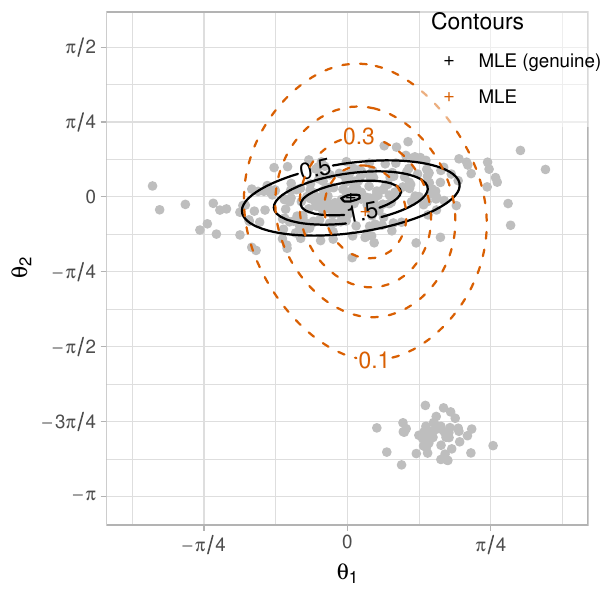}
	\includegraphics[width=0.45\textwidth]{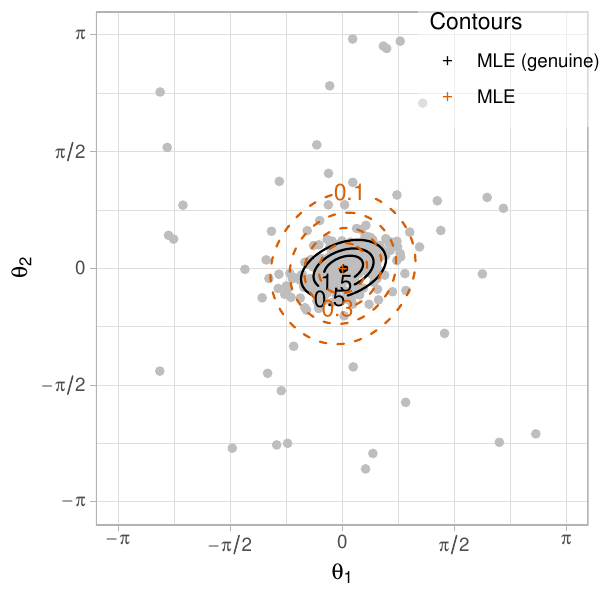}
	\caption{Fitted density contours for the two contaminated datasets. Contours based on approximate maximum likelihood estimates are superimposed on the data. Left: the outliers are clustered. Right: outliers are scattered.}
	\label{fig: mle-contam}
\end{figure}

Outliers can have a severe effect on the estimation process of the parameters of the von Mises sine distribution in \eqref{eq:dvmsin}. Consider the data described in Section \ref{sec:2}. Then, we add $50$ outliers that clearly deviate from the pattern shared by the genuine data: outliers are clustered in the top row panels of Figure \ref{fig: torus-contam}, whereas they are scattered in the bottom row panels of Figure \ref{fig: torus-contam}. In the former scenario,  we note that contamination only concerns one dimension, but it affects both dimensions in the second case. Figure \ref{fig: mle-contam} shows the fitted model density in the presence of contamination for both considered outliers' constellations: contour lines are completely different from the uncontaminated scenario, they are both shifted and rotated. This happens because the data are not homogeneous.

With this aim in mind, in the presence of contaminated data, it is suggested to replace the estimates in (\ref{esteq}) with their weighted counterparts 
\begin{align}
	\label{westeq}
	\tilde{\mu}_j & = \mathrm{atan2} (\tilde{U}_j, \tilde{V}_j) \ , ~j=1, \dots, p \ , \nonumber \\
	\tilde \Sigma_{jj} & = \frac{2}{\sum_{i=1}^n w_i} \sum_{i=1}^n w_i \left(1-\cos(\theta^{(i)}_j-\tilde{\mu}_j)\right) \ , ~j=1, \dots, p \ , \\
	\tilde \Sigma_{jl} & = \frac{1}{\sum_{i=1}^n w_i}\sum_{i=1}^n w_i \sin(\theta^{(i)}_j-\tilde{\mu}_j)\sin(\theta^{(i)}_l-\tilde{\mu}_l) \ , ~l \neq j,~j=1, \dots, p \ , \nonumber  
\end{align}
where the weights are as in \eqref{weight} and
$\tilde{U}_j, \tilde{V}_j$ are the weighted sample means of $\cos(\theta_j)$ and $\sin(\theta_j)$ respectively. 
We point out that detecting outliers on the torus using the Pearson residual function \eqref{pearson} is particularly appealing. In fact, defining and measuring outlyingness using other notions of (geometric) distance on the torus is not straightforward~\citep{mardia2012statistics}.

\begin{figure}[!ht]
	\centering
	\includegraphics[scale=0.6]{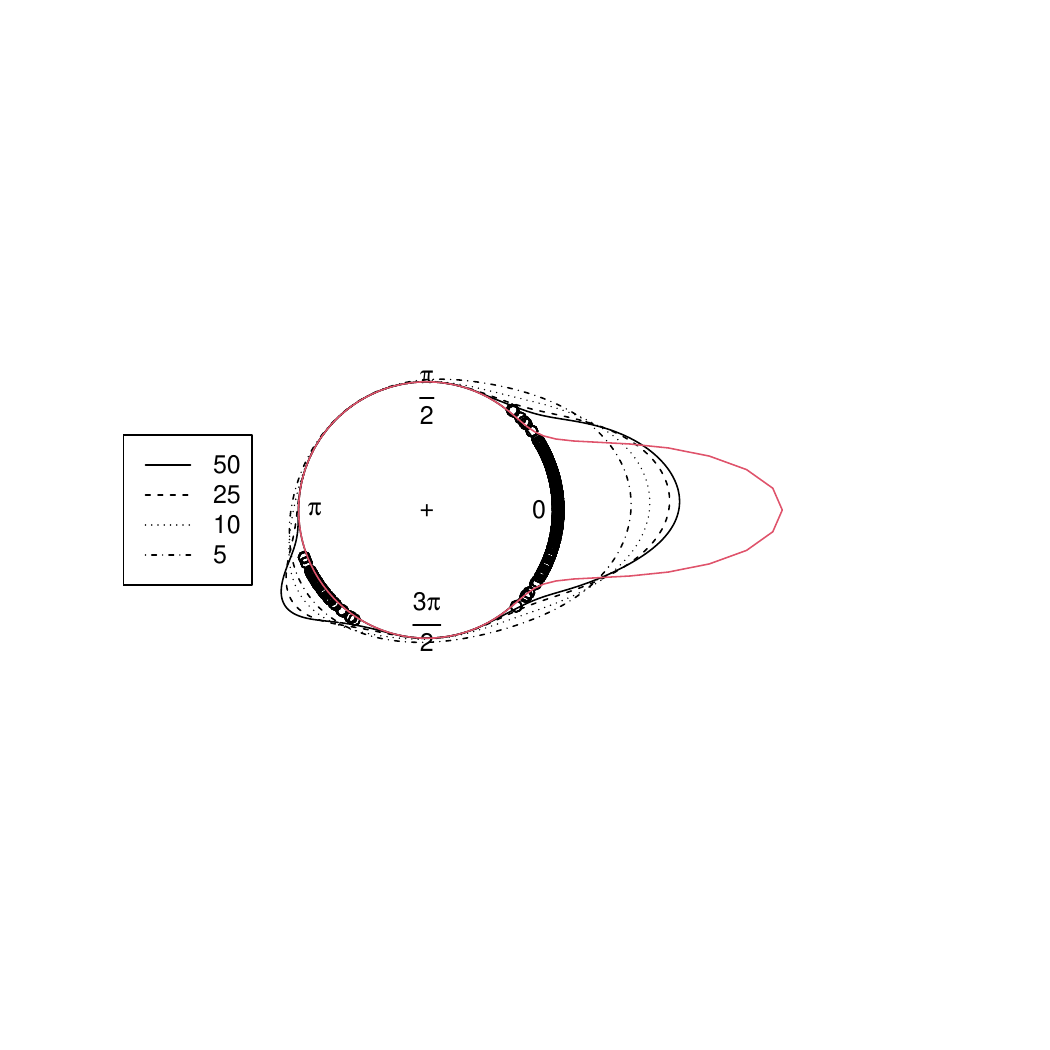}
	\vspace{-3cm}
	\caption{Fitted marginal densities. Sample from the bivariate von Mises distribution with clustered outliers. Marginal densities for the component $\theta_2$ are shown for different bandwidths. The true model is indicated by the red line.}
	\label{fig:k}
\end{figure}

The first key ingredient for weighted likelihood estimation on the torus is a non-parametric estimate $\hat{f}_n(\vect{\theta})$. Here, we restrict our attention to kernel density estimates of the form \citep{bai1988kernel, jung2021clustering, di2011kernel}
$$
\hat{f}(\vect{\theta}) = \frac{1}{n}\sum_{i=1}^nK \left(\vect{k}^*\cos(\vect{\theta}-\vect{\theta}^{(i)})\right)
$$
where $K\left(\vect{k}^*\cos(\vect{\theta})\right)=\prod\limits_{j=1}^p c(k_j^*)K\left(k_j^*\cos(\theta_j)\right)$ is the $p-$fold product of 1-dimensional toroidal kernels.
In particular, it is possible to consider component-wise von Mises kernels indexed by the same concentration bandwidth parameter $k^*$ \citep{agostinelli2007robust}. A von Mises kernel function is obtained for $K(t)=\exp(t)$ with normalizing constant $c(k^*)=\left(2\pi I_0(k^*)\right)^{-1}$, where $I_0(x)$ is the modified Bessel function.
The effect of the bandwidth concentration parameter is shown in Figure \ref{fig:k}. We notice that the larger $k^*$ the smaller the smoothing. Then, large values of the concentration bandwidth can make the non-parametric circular fitted density $\hat{f}(\vect{\theta})$ sensitive to the presence of contamination in the data, whereas, on the contrary, smaller values can mask outliers due to over-smoothing.

\begin{figure}[!ht]
	\centering
	\includegraphics[width=0.45\textwidth]{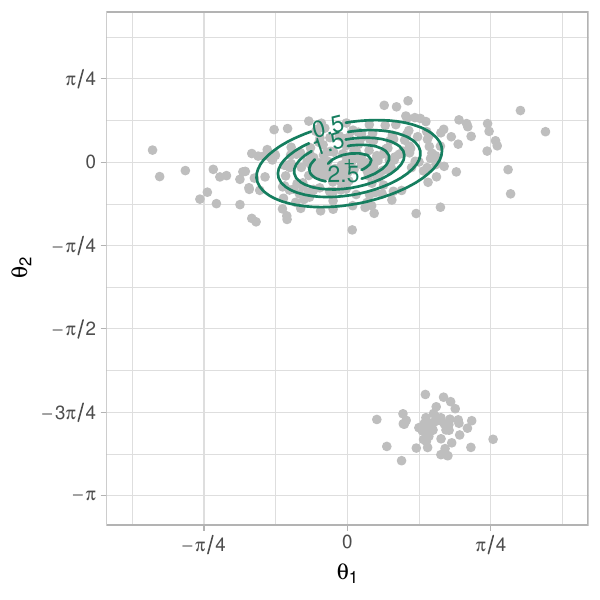}
	\includegraphics[width=0.45\textwidth]{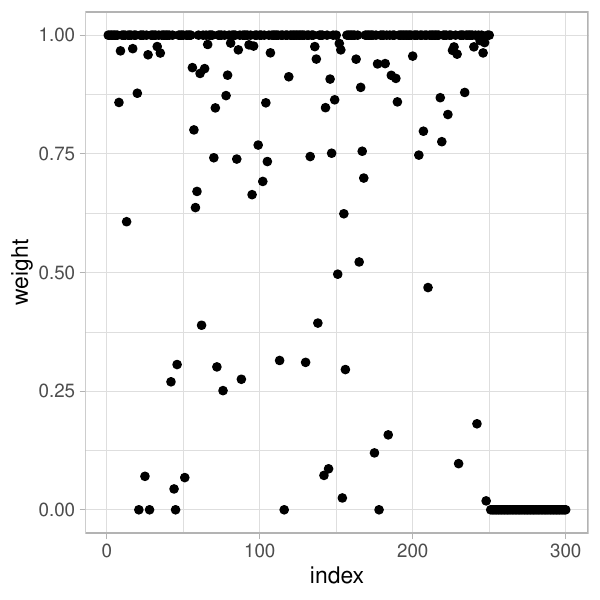}\\
	\includegraphics[width=0.45\textwidth]{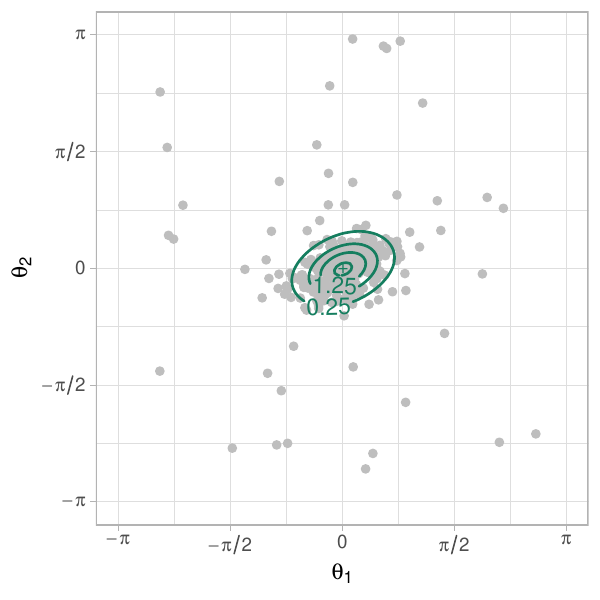}
	\includegraphics[width=0.45\textwidth]{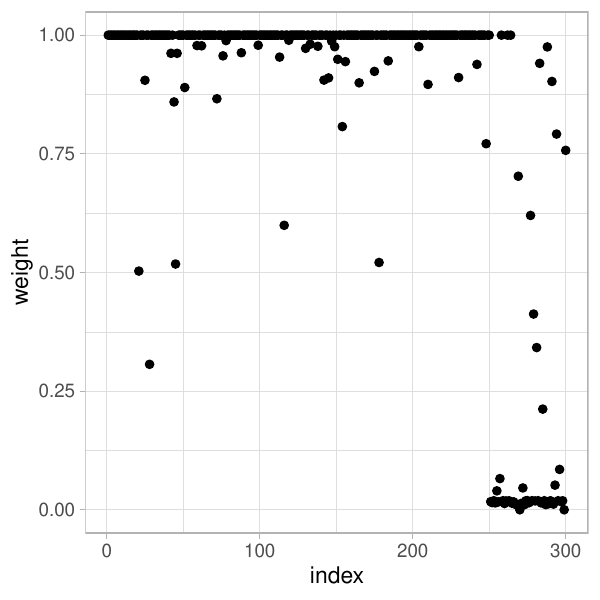}
	\caption{WLE fitted density contours and weights. The $n = 250$ genuine data points are contaminated by 50 clustered outliers (top row) and 50 scattered outliers (bottom row). Left: WLE fitted density contours. Right: weights.}
	\label{fig:wle-cont-ws}
\end{figure}

However, the von Mises distribution is not closed under convolution and the convoluted smoothed density $\hat{m}(\vect{\theta})$ cannot be expressed in closed form, being proportional to
$$
\int_{\mathbb{T}^p} \exp\left[\vect{\kappa}^* \cos(\vect{\theta}-\vect{t}) + \vect{\kappa}^\top\cos(\vect{\theta}-\vect{\mu}) + \frac{1}{2}\sin(\vect{\theta}-\vect{\mu})^\top \Lambda \sin(\vect{\theta}-\vect{\mu})\right] \ d\vect{t} \ .
$$
Then, Pearson residuals are evaluated comparing the fitted kernel density to the actual non-smoothed model as in \eqref{eq:residualfs}. This approach is feasible for data on the torus as long as the model density remains strictly positive over the entire support. This property prevents the occurrence of very small (near-zero) densities that could otherwise destabilize the denominator of the Pearson residuals \citep{agostinelli2019weighted, agostinelli2024weighted}.
The left panels of Figure \ref{fig:wle-cont-ws} show the contours of the fitted model density over the contaminated sample data obtained from WLE based on a GKL RAF (here $k^*=25$) for clustered (top) and scattered (bottom) outliers. In both scenario, the fit provided by the WLE exhibits negligible differences from that given by the MLE on the uncontaminated sample. Furthermore, the inspection of the weights in the corresponding right panels of Figure \ref{fig:wle-cont-ws} allows to clearly discriminate between the genuine data and the outliers.

Bandwidth selection should be pursued according to a monitoring strategy \citep{agostinelli2018discussion, greco2020weighted, greco2020weighted1}. It is suggested
to run the procedure for different values of the concentration bandwidth $k^*$ and monitor the behavior of estimates and/or weights (or a summary of them given by the empirical downweighting level) as it varies in a reasonable range. The resulting fit becomes more robust as $k^*$ grows. Then, the monitoring is expected to show a possible transition from a non-robust to a robust fit: some weights start being small from some concentration bandwidth value on, or the empirical downweighting level shows an increasing trend before its trajectory reaches a plateau.
One can select a concentration bandwidth at least equal to the value where this transition occurs. This strategy is illustrated in Section \ref{sec:6}.

The iterative algorithm to solve the WLEE in \eqref{westeq} can be initialized using subsampling.
The initial subsample is supposed to be free of outliers. Its size is therefore chosen to be small, to increase the probability of selecting an uncontaminated subset, yet sufficiently large to allow the actual estimation of the unknown parameters.
The mean parameter $\vect{\mu}$ is initialized using the circular sample mean, $\bar{\vect{\theta}}$.
Initial diagonal elements of $\Sigma$ can be obtained as $\Sigma^{(0)}_{rr}=-2\log(\hat{\rho}_r)$, where $\hat{\rho}_r$ is the sample mean resultant length, whereas its off-diagonal elements are given by 
$\Sigma^{(0)}_{rs}=r_{\text{JS}}(\vect{\theta}_r, \vect{\theta}_s) \sqrt{\Sigma_{rr}^{(0)} \Sigma_{ss}^{(0)}}$, where 
$$
r_{\text{JS}}(\vect{\theta}_r, \vect{\theta}_s) =\frac{\sum\limits_{i=1}^n \sin \left(\theta_r^{(i)}-\bar{{\theta}}_r\right) \sin \left(\theta_s^{(i)}-\bar{\theta}_s\right)}{\sqrt{\left[\sum\limits_{i=1}^n\left(\sin \left(\theta_r^{(i)}-\bar{{\theta}}_r\right)\right)^2\right] \left[\sum\limits_{i=1}^n\left(\sin \left(\theta_s^{(i)}-\bar{{\theta}}_s\right)\right)^2\right]}} 
$$
is the circular correlation coefficient, for $r, s = 1, \ldots, p$ and $r\neq s$, defined by~\citet{JammalamadakaSenGupta2001}. 

The approximate MLE evaluated on the subsample is an alternative method for initialization.
A different strategy to obtain a starting outliers free subsample consists in fitting an oversmoothed circular kernel density first, and then select a fixed fraction (up to half of the data) of points with the largest densities.
However, it is suggested to run the algorithm from several starting points.
The {\it best} solution can be selected by minimizing the probability to observe a small Pearson residual. This probability can be fitted using random draws from the fitted model \citep{agostinelli2019weighted, saraceno2021robust}.

Under no contamination, consistency and asymptotic normality of the WLE for the parameters of the von Mises distribution follow from the general
results established in \citep{agostinelli2019weighted, agostinelli2024weighted}. 
The assumptions required are easily verified for the von Mises, and they can be generalized as follows
\begin{enumerate}
	\item the kernel $K(\vect{k}^*\cos(\vect{\theta}))$ is of bounded variation;
	\item the model is correctly specified, that is, there exists $\vect{\tau}_0 \in T$ such that $f(\vect{\theta}) = m(\vect{\theta}; \vect{\tau}_0)$ a.s., with $\vect{\tau}=(\vect{\mu}, \vect{\kappa}, \Lambda)$;
	% \item $f^\circ(\vect{\theta}) = m^\circ(\vect{\theta}; \vect{\tau}_0)$
	\item the model density is positive over the support, that is, there exists $B > 0$ such that $\sup_{\vect{\theta}, \vect{\tau}} m(\vect{\theta}; \vect{\tau}) \geq B$;
	\item the RAF satisfies conditions $A(0)=0$, $A^\prime(0)=1$ and $A^{\prime\prime}(\delta)$ is a bounded and continuous function w.r.t. $\delta$;
	\item
	$\Psi(\vect{\theta};\vect{\tau}) = w\left(\delta(\vect{\theta};\vect{\tau})\right) u(\vect{\theta}; \vect{\tau})$ is differentiable, the Jacobian matrix $\dot{\Psi}(\vect{\theta};\vect{\tau})$, with elements $ij$ given by $\partial \Psi_i/\partial \tau_j$, is positive definite and $\mathbb E_{\vect{\tau}_0}(\dot{\Psi}(\vect{\theta};\vect{\tau}))$ is finite $ \forall(\vect{\theta},\vect{\tau})$.
\end{enumerate}

\section{Numerical studies}
\label{sec:5}

In this section, we investigate the finite-sample performance of the proposed weighted likelihood estimator (WLE) for the parameters of a multivariate von Mises sine distribution. The study considers dimensions $p=2$ and $p=5$, with sample size fixed at $n=300$, and is based on $5{,}000$ Monte Carlo replications.

In the bivariate case, data are generated according to the same mechanism as in Section~\ref{sec:4}, with concentration parameters $\vect{\kappa} = (10, 20)$ and dependence parameter $\lambda = 5$.  
For the five-dimensional case, observations are constructed using independent blocks: two independent bivariate von Mises sine distributions generated as above are combined with an independent univariate von Mises distribution with concentration parameter $\kappa_0 = 20$. In all cases, the marginal circular means are set to zero.

Contaminated samples are obtained by replacing a fraction $\lfloor n\epsilon \rfloor$ of the genuine observations with outliers, using a contamination rate $\epsilon = 10\%$. Both clustered and scattered contamination schemes are considered, as in the examples of Section~\ref{sec:4}.  
When $p=5$, contamination affects only the first two dimensions. In both dimensions, outliers are generated from a bivariate normal distribution with:
(i) $\vect{\mu} = c(0.5, -2.5)$ and covariance $\sigma_{ii} = 0.002$, $\sigma_{ij} = 0$ for \emph{clustered} outliers, and  
(ii) $\vect{\mu} = c(0, 0)$ and covariance $\sigma_{ii} = \pi$, $\sigma_{ij} = 0$ for \emph{scattered} outliers, where $i,j = 1,2$ and $i \neq j$.  
The uncontaminated case is also included to assess the efficiency loss of the WLE under the genuine model.

Throughout, we adopt a GKL RAF with tuning constants $k^* = 200$ in the bivariate case and $k^* = 25$ in the five-dimensional case. The WLE is initialized using an oversmoothed pilot circular density estimate, obtained with a concentration bandwidth $k^*=2$; this choice was guided by preliminary pilot runs.

The accuracy and precision of the WLE and the MLE are assessed using the square root mean angle separation 
$$
\sqrt{AS(\hat{\vect{\mu}})}=\sqrt{\frac{1}{p}\sum_{j=1}^p(1-\cos(\hat\mu_j))}
$$ 
and the root mean squared errors (RMSE) 
$$
\|\hat{\vect{\kappa}} - \vect{\kappa}\|_2
\quad \text{and} \quad
\|\hat{\vect{\lambda}} - \vect{\lambda}\|_2,
$$
where $\vect{\lambda}=(\lambda_{12}, \dots, \lambda_{(p-1)p})^\top$ collects the upper-triangular elements of the dependence matrix $\Lambda$. 

In addition to the proposed WLE based on the von Mises sine model (denoted VM--VM), we also consider two alternative WLEs. The first is a WLE for $(\vect{\mu}, \Sigma)$ under a Wrapped Normal (WN) model, computed via an EM-type algorithm with weights evaluated on the torus using a WN kernel~\citep{saraceno2021robust, agostinelli2024weighted}.  
To facilitate a fair comparison, the second WLE  is obtained under the von Mises sine model, but the weights are computed using the WN kernel torus density estimate (denoted VM--WN). In both cases, a diagonal bandwidth matrix $\vect{H} = h\,\mathrm{I_d}$ indexes the WN kernel, with $h \approx 1/k^*$.

% ---
% no contamination
% ---

\begin{table}[t]
	\scriptsize
	\begin{subtable}[h]{0.42\textwidth}
		\centering
		\begin{tabular}{lrrr}
		\hline
		& eff (VM-VM) & eff (VM-WN) & eff (WN) \\
		\hline
		$\hat{\vect{\mu}}$ & 0.9933 & 0.9933 & 1.0028 \\
		$\hat{\vect{\kappa}}$ & 0.2894 & 0.2893 & 0.9218 \\
		$\hat{\lambda}$ & 0.9004 & 0.9004 & 1.2172 \\
		\hline
		\end{tabular}
		\caption{bivariate case}
	\end{subtable}
	\qquad \qquad
	\begin{subtable}[h]{0.42\textwidth}
		\centering
		\begin{tabular}{lrrr}
		\hline
		& eff (VM-VM) & eff (VM-WN) & eff (WN) \\
		\hline
		$\hat{\vect{\mu}}$ & 0.9999 & 0.9979 & 0.9992 \\
		$\hat{\vect{\kappa}}$ & 0.3443 & 0.3908 & 0.9662 \\
		$\hat{\vect{\lambda}}$ & 0.7434 & 0.7690 & 0.9372 \\
		\hline
		\end{tabular}
		\caption{five-dimensional case}
	\end{subtable}
\caption{Monte Carlo simulations. Efficiency of the considered WLEs is evaluated taking the ratio between the Mean Squared Error of the MLE and of the WLE.}
	\label{tab:mse-rt}
\end{table}

\begin{figure}[!ht]
	\centering
	\includegraphics[width=0.9\textwidth]{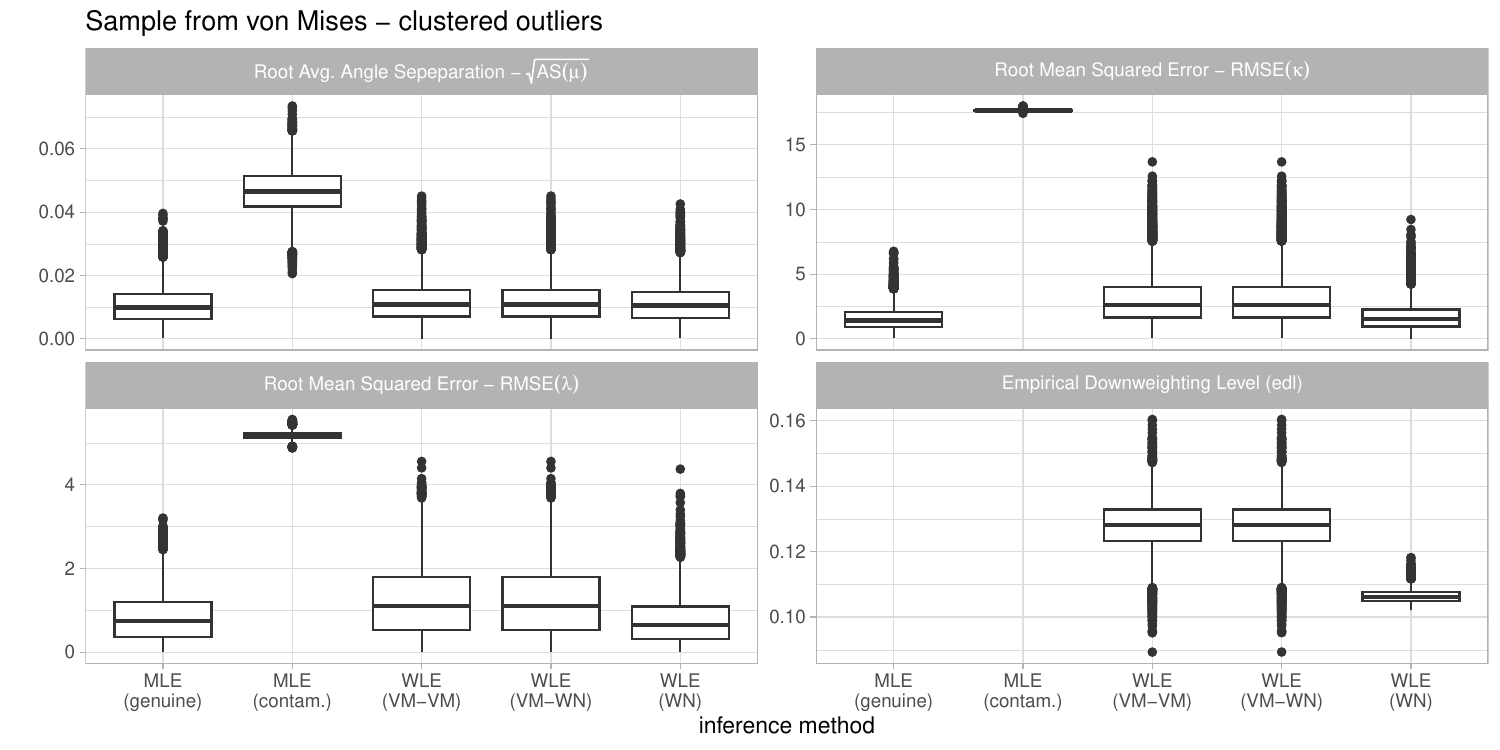}
	\caption{Monte Carlo simulation (bivariate). Empirical distributions of the estimation errors for parameters $\vect{\mu}$, $\vect{\kappa}$, and $\lambda$, together with the empirical downweighting level (edl). Different estimation methods are displayed along the $x-$axis. Results are based on samples of size $n=300$ from a bivariate von Mises sine distribution with clustered outliers.}
	\label{fig:sim2d-acc-vm-gr}
\end{figure}

\begin{table}[h]
\begin{tabular}{l|r|r|r|r|r|}
\hline
& $\sqrt{AS(\hat{\vect{\mu}})}$ & RMSE($\hat{\vect{\kappa}}$) & RMSE($\hat{\lambda}$) & edl & WLRT  \\ \hline
MLE (genuine)& 0.0106 & 1.5761 & 0.8245 &  --    & 5.4883 \\
MLE (cont.)  & 0.0466 & 17.6526 & 5.1799 &  --   & 1614.0887 \\
WLE (VM-VM)  & 0.0118 & 3.0853 & 1.2310 & 0.1279 & 12.9497 \\
WLE (VM-WN)  & 0.0118 & 3.0846 & 1.2307 & 0.1279 & 12.9440 \\
WLE (WN)     & 0.0113 & 1.7073 & 0.7609 & 0.1064 & 5.2043 \\ \hline
\end{tabular}
\caption{Monte Carlo simulation (bivariate). Average estimation errors for parameters $\vect{\mu}$, $\vect{\kappa}$, and $\lambda$, together with the average empirical downweighting level (edl), for different estimation methods. Results are based on samples of size $n=300$ from a bivariate von Mises sine distribution with clustered outliers.}
\label{tab:mc2d-avgs-1}
\end{table}

\begin{figure}[ht]
	\centering
	\includegraphics[width=0.9\textwidth]{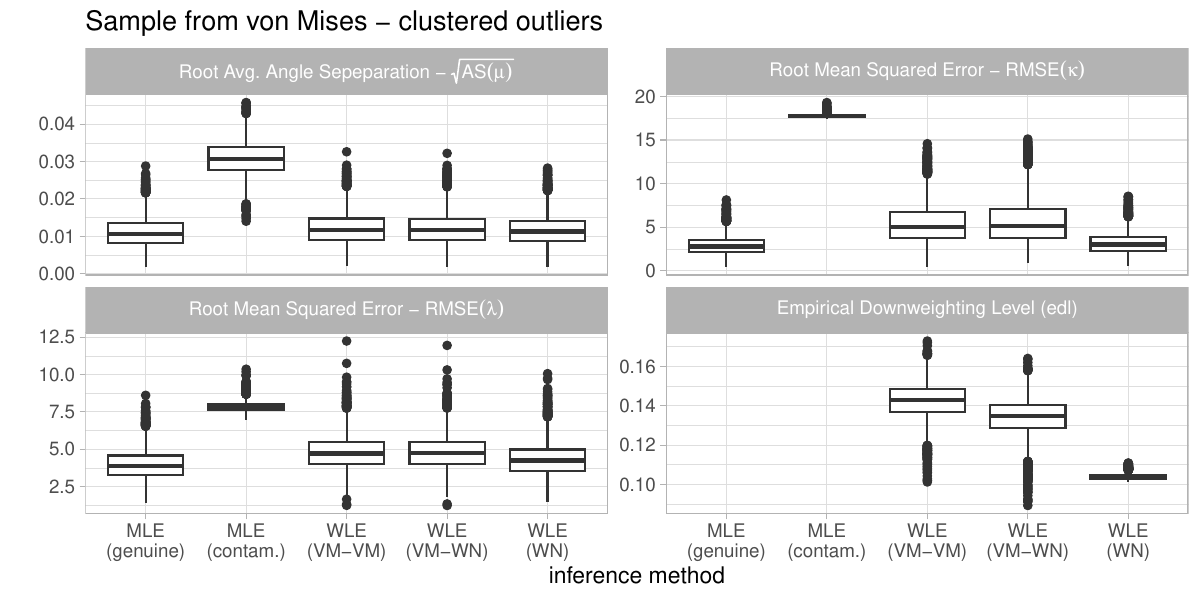}
	\caption{Monte Carlo simulation (five-dimensional). Empirical distributions of the estimation errors for parameters $\vect{\mu}$, $\vect{\kappa}$, and $\lambda$, together with the empirical downweighting level (edl). Different estimation methods are displayed along the $x-$axis. Results are based on samples of size $n=300$ from a five-dimensional von Mises sine distribution with clustered outliers.}
	\label{fig:sim5d-acc-vm-gr}
\end{figure}

To quantify the efficiency loss of the proposed WLEs relative to the MLE, we compute the ratio between the mean squared error of the MLE and those of the WLEs. The results are reported in Table~\ref{tab:mse-rt}. 
It can be observed that the proposed WLE incurs a substantial loss of efficiency in the estimation of the concentration parameters. The WLE based on the von Mises sine model generally underestimates the concentration parameter.  

Figure~\ref{fig:sim2d-acc-vm-gr} displays the results for the bivariate setting with clustered outliers, while average values are given in Table~\ref{tab:mc2d-avgs-1}. 
The square root average angle separation is close to zero for all three WLEs as well as for the MLE without contamination, with no noticeable differences in their empirical distributions. In contrast, when contamination occurs, the angle separation for the MLE exhibits a markedly different behavior, with its distribution shifting upward.

As expected, no appreciable differences arise between the WLE for the von Mises sine model when weights are computed using either a von Mises (VM--VM) or a Wrapped Normal (VM--WN) kernel, provided that the bandwidths are selected consistently. 
Interestingly, under the bivariate Wrapped Normal model, the WN WLE yields, on average, slightly more precise estimates for $\lambda$ and $\vect{\kappa}$, along with an empirical downweighting level closer to the true contamination rate $\epsilon = 0.1$. 
This behavior can be attributed to the fact that, under high concentration, the von Mises distribution is well approximated by a normal distribution~\citep{mardia2012statistics}. Moreover, the Wrapped Normal model enjoys more favorable analytical properties than the multivariate von Mises distribution and, above all, the maximum likelihood estimators for the WN distribution are exact, while the estimators \eqref{esteq} are only approximate. 
In fact, in~\cite{mardia2012mixtures} the authors shown that the approximated MLEs~\eqref{esteq} are close to the numerically computed exact MLEs for particular combinations of parameters, while otherwise they underestimate the true parameters. In the Appendix~\ref{secA1} we show an example where the proposed WLE based on the von Mises kernel and model yields estimates for $\vect{\kappa}$ which are even more accurate than the MLEs for the genuine data.

Analogous results for the five-dimensional simulations are shown in Figure~\ref{fig:sim5d-acc-vm-gr}, while Figures~\ref{fig:sim2d-acc-vm-sc}--\ref{fig:sim5d-acc-vm-sc} in the Appendix~\ref{secA1} display the results for the scattered-outliers scenario.

\begin{figure}[ht]
	\centering
	\includegraphics[width=0.9\textwidth]{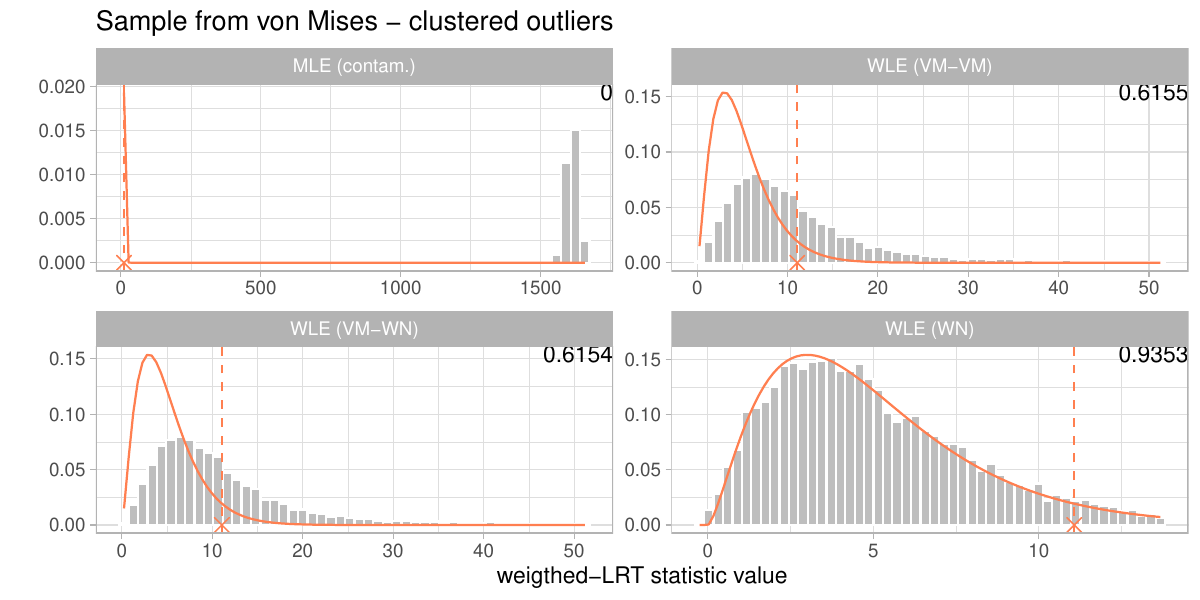}
	\caption{Monte Carlo simulations (bivariate). Sampling distributions for the WLRT (LRT for MLE under contamination) statistic and the $\chi^2-$ density superimosed. Empirical coverage at level 0.95 printed on the top-right of each subplot. Results are based on samples of size $n=300$ from a bivariate von Mises sine distribution with clustered outliers.}
	\label{fig:wlrt-gr}
\end{figure}

\begin{figure}[!ht]
	\centering
	\includegraphics[width=0.9\textwidth]{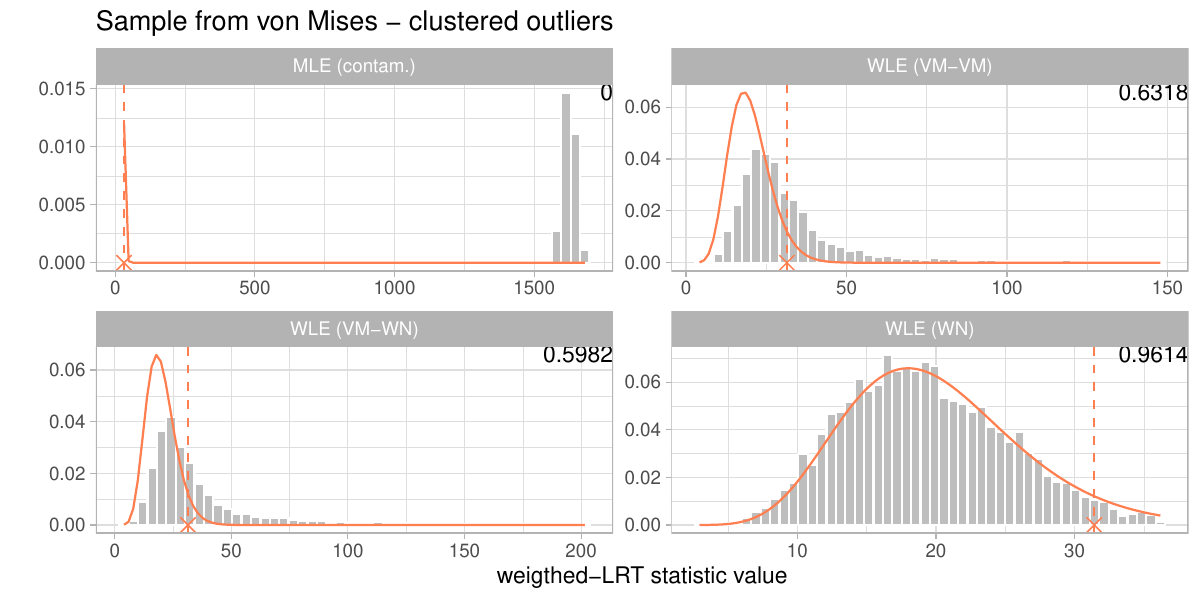}
	\caption{Monte Carlo simulations (five-dimensional). Sampling distributions for the WLRT (LRT for MLE under contamination) statistic and the $\chi^2-$ density superimosed. Empirical coverage at level 0.95 printed on the top-right of each subplot. Results are based on samples of size $n=300$ from a five-dimensional von Mises sine distribution with clustered outliers.}
	\label{fig:wlrt-5d-gr}
\end{figure}

We also investigate the behavior of the weighted likelihood ratio test \citep[WLRT,][]{agostinelli2001test} in comparison with the classical likelihood ratio test (LRT) under contamination. The WLRT statistic for testing the null hypothesis $H_0:\vect{\tau}=\vect{\tau}_0$ is given by
\begin{equation*}
	\Lambda_{W}(\vect{\tau}_0)
	= -2 \sum_i w_i \left\{ \ell(\vect{\theta}^{(i)}; \vect{\tau}_0)
	- \ell(\vect{\theta}^{(i)}; \hat{\vect{\tau}}_{W}) \right\},
\end{equation*}
and is known to follow an asymptotic $\chi^2_q$ distribution under the null hypothesis, with $q = \mathrm{dim}(T)$.
Figures~\ref{fig:wlrt-gr}--\ref{fig:wlrt-5d-gr} show the sampling distributions of the WLRT for the bivariate and five-dimensional simulated circular data with clustered outliers, together with the empirical coverages for a $0.95$ nominal level. 
Consistently with the previous results, the WLRT provides improved inferential performance under contamination compared to the classical LRT. 
Once again, the WN WLE outperforms the VM-based WLEs, yielding empirical coverage levels closer to the nominal value. Figures~\ref{fig:wlrt-sc}-\ref{fig:wlrt-5d-sc} show similar results for scattered outliers. 
Tables~\ref{tab:coverage-vm}-\ref{tab:coverage-vm-5} 
in the Appendix~\ref{secA1} collect together the coverage values in the different contamination scenarios.

\section{Empirical applications}
\label{sec:6}

In this section, we illustrate the proposed procedure using two real data sets. The first is a bivariate data set concerning protein torsion angles, while the second is a five-dimensional data set involving wind direction.

\subsection{TIM8 protein torsion angles}
\begin{figure}[ht]
	\centering
	\includegraphics[width=0.45\textwidth]{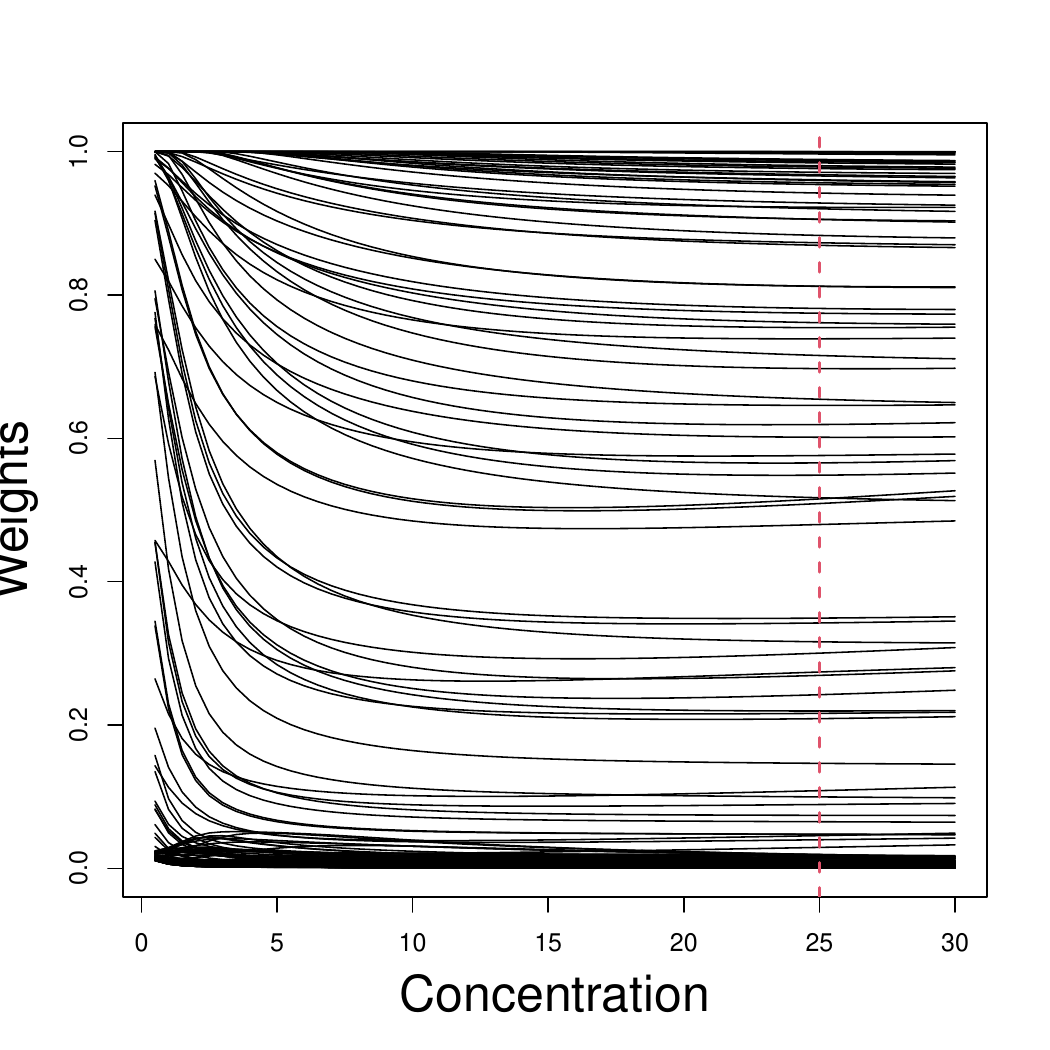}
	\includegraphics[width=0.45\textwidth]{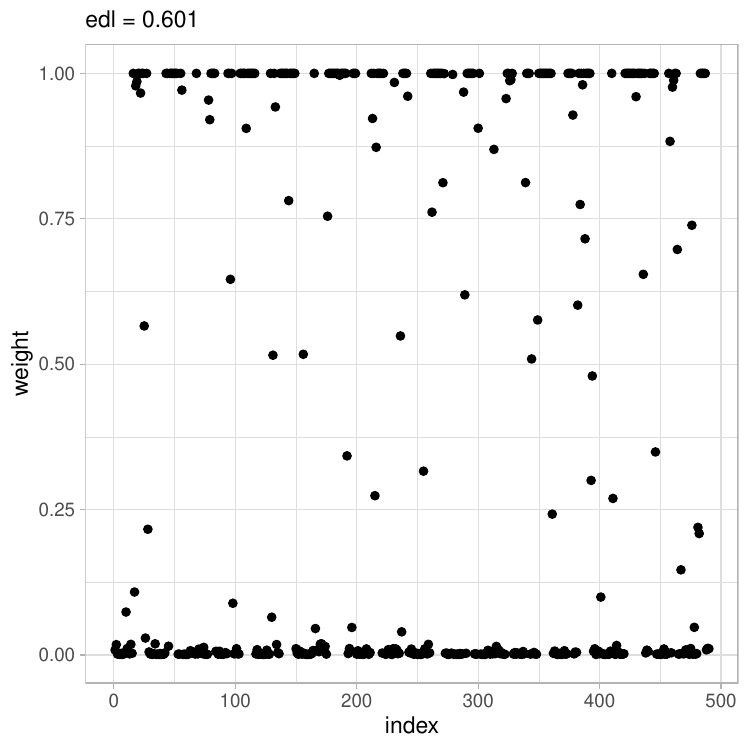}
	\caption{TIM8 protein data. Left: Monitoring plot of the weights as a function of the concentration bandwidth. The red vertical line indicates $k^*=25$. Right: Weights at the selected bandwidth. The weights are adjusted according to the GKL RAF.}
	\label{fig:5}
\end{figure}

\begin{figure}[ht]
	\centering
	\includegraphics[width=0.45\textwidth]{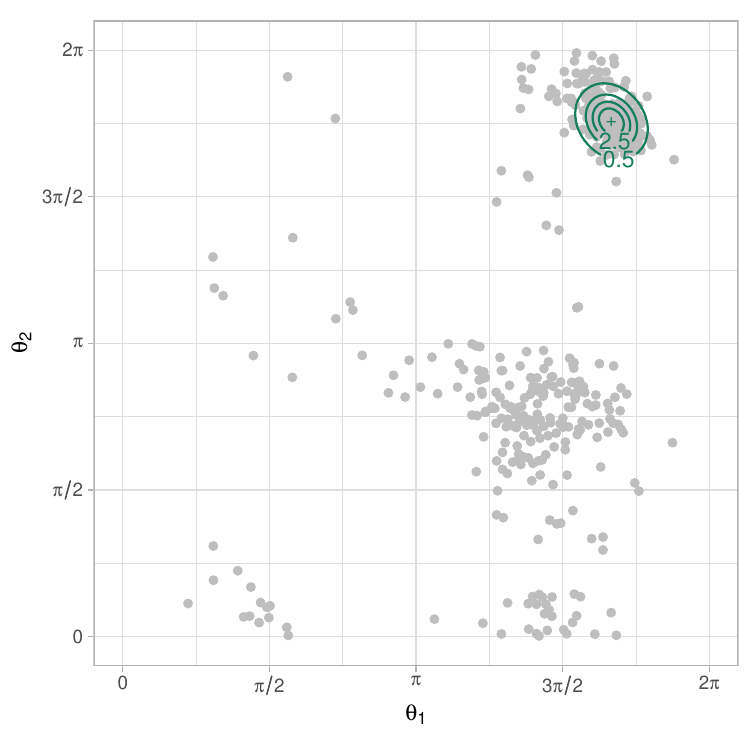}
	\includegraphics[width=0.45\textwidth]{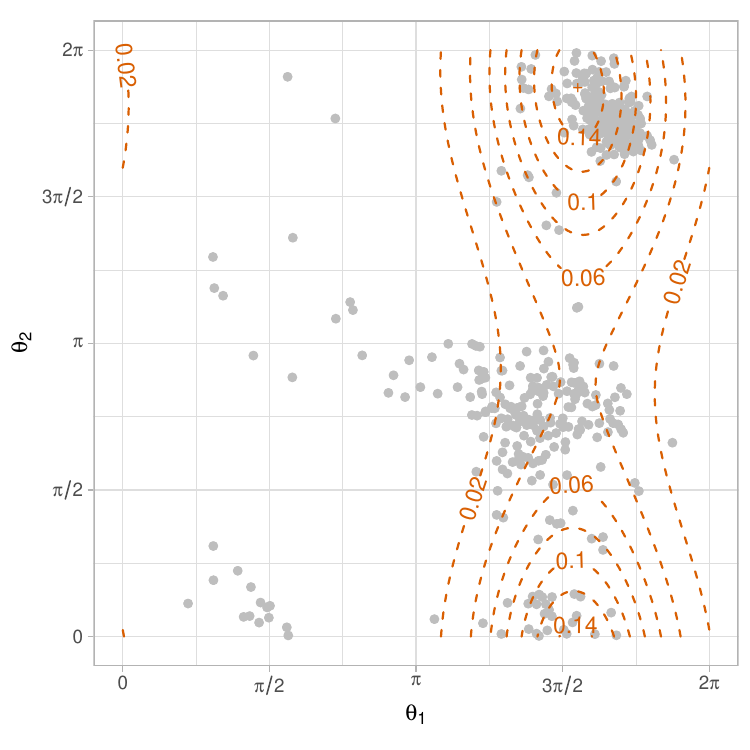}
	\caption{TIM8 protein data. Left: fitted density contours from WLE. Right: fitted density contours from MLE. The point estimate is denoted by the symbol +.}
	\label{fig:6}
\end{figure}

The 8TIM data set from the \texttt{R} package \texttt{BAMBI} \citep{bambi} consists of $n=490$ backbone torsion angle pairs $(\phi, \psi)$ for the protein 8TIM. The data exhibit a multimodal, clustered structure.
Maximum likelihood estimation is not able to unveil such structures. In contrasts, the WLE analysis, based on a GKL RAF function (here $k^*=25$), provides strong evidence of the presence of several otherwise undetectable clusters, revealed through the inspection of the weights, which are shown in the right panel of Figure~\ref{fig:5}. 
The empirical downweighting level is about $0.6$, meaning that more than half of the data pertain to different random mechanisms.
Initialization is based on an oversmoothed kernel density on the torus.
The left panel of Figure~\ref{fig:5} displays the monitoring plot of the weights as the bandwidth concentration parameter varies. The trajectories describe a smooth transition from a non-robust to a robust fit as the concentration grows. We notice that many trajectories decline as we move to the right of the plot, meaning that the corresponding observations exhibited some deviations from the model that were to be taken properly into account.
Actually, the fitted model density by WLE is mainly concentrated over one group of angles, as shown in the left panel of Figure \ref{fig:6}. The reader is pointed to \cite{agostinelli2024weighted} for a similar result under different model assumptions. The right panel of Figure \ref{fig:6} displays the maximum likelihood fitted model. 
The contour density lines are meant to be wrapped onto a torus: one should topologically glue both pairs of opposite edges together with no twists, hence obtaining the torus $\mathbb{T}^2$.
As it is shown in Table \ref{tab:1} there is a substantial difference between the parameter estimates according to the two estimated models, especially in the dependence structure, that is summarized by the parameters $k_1, k_2$, and $\lambda$.

\begin{table}[t]
	\centering
	\begin{tabular}{c|ccccc}
		\hline
		& $\phi$ & $\psi$ & $k_1$ & $k_2$ & $\lambda$\\
		\hline
		MLE & 4.87 & 5.87 & 1.28 & 0.60 & 0.06 \\
		WLE & 5.21 & 5.55 & 8.81 & 8.48 & -1.13 \\
		\hline
	\end{tabular}
	\caption{TIM8 protein data. Parameter estimates from WLE and MLE.}
	\label{tab:1}
\end{table}

\subsection{Col De La Roa wind directions}

\begin{figure}[!ht]
	\centering
	\includegraphics[width=0.31\textwidth]{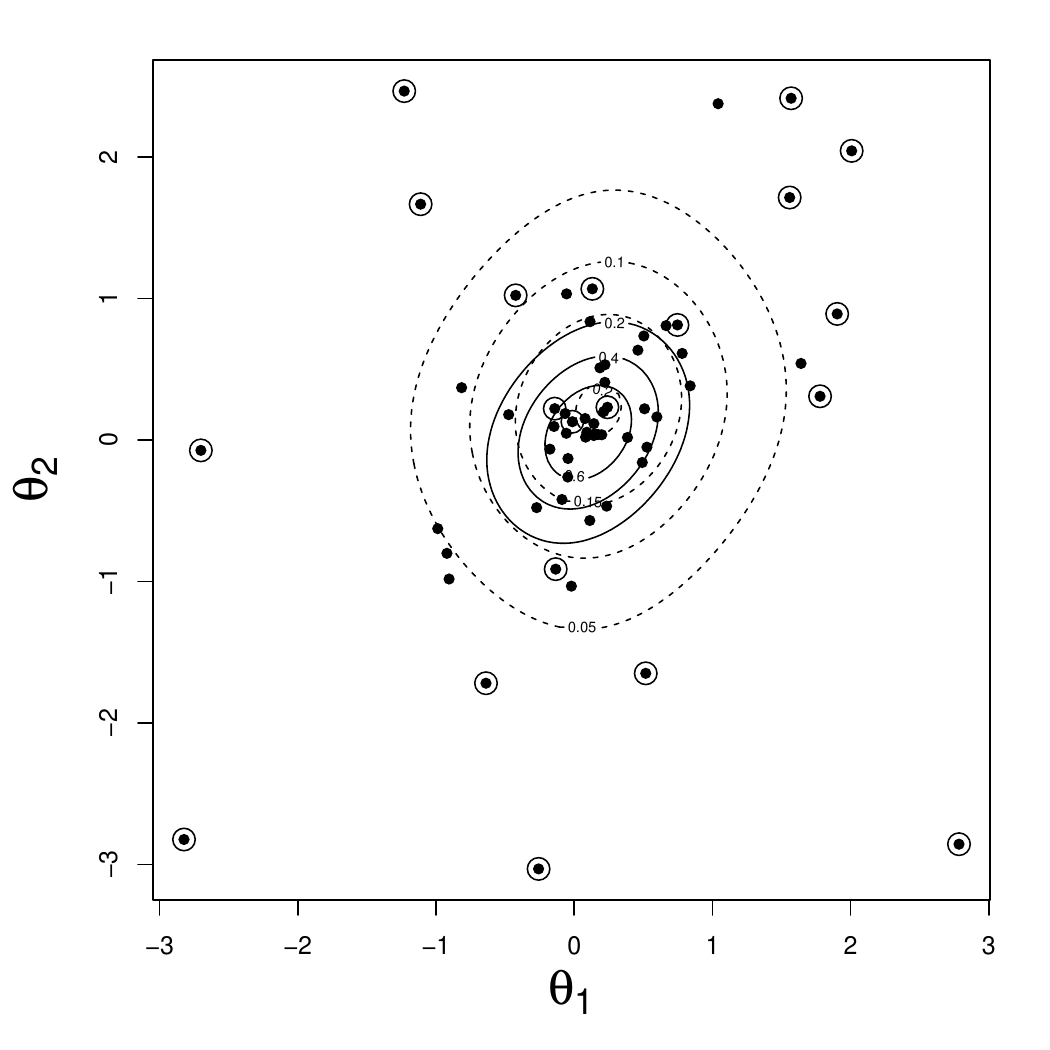}
	\includegraphics[width=0.31\textwidth]{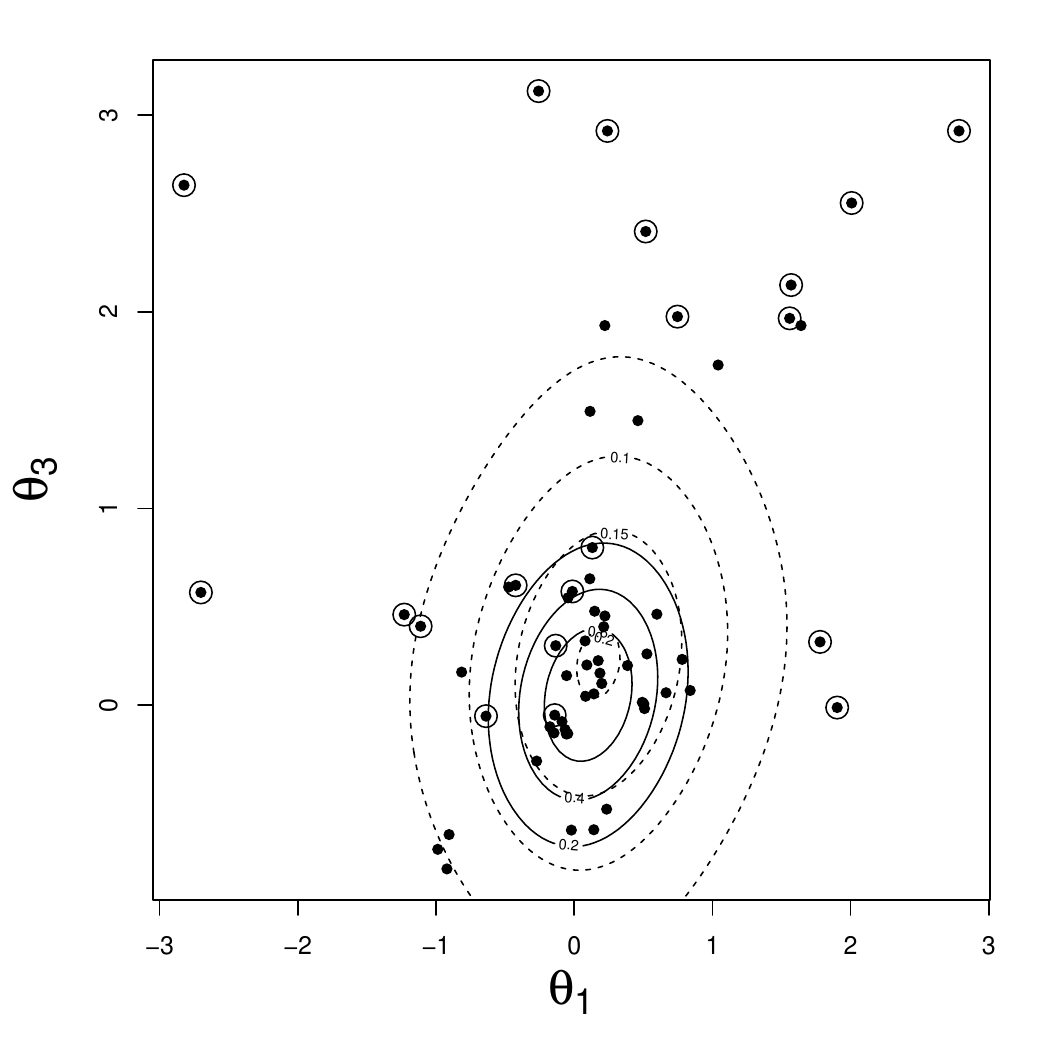}
	\includegraphics[width=0.31\textwidth]{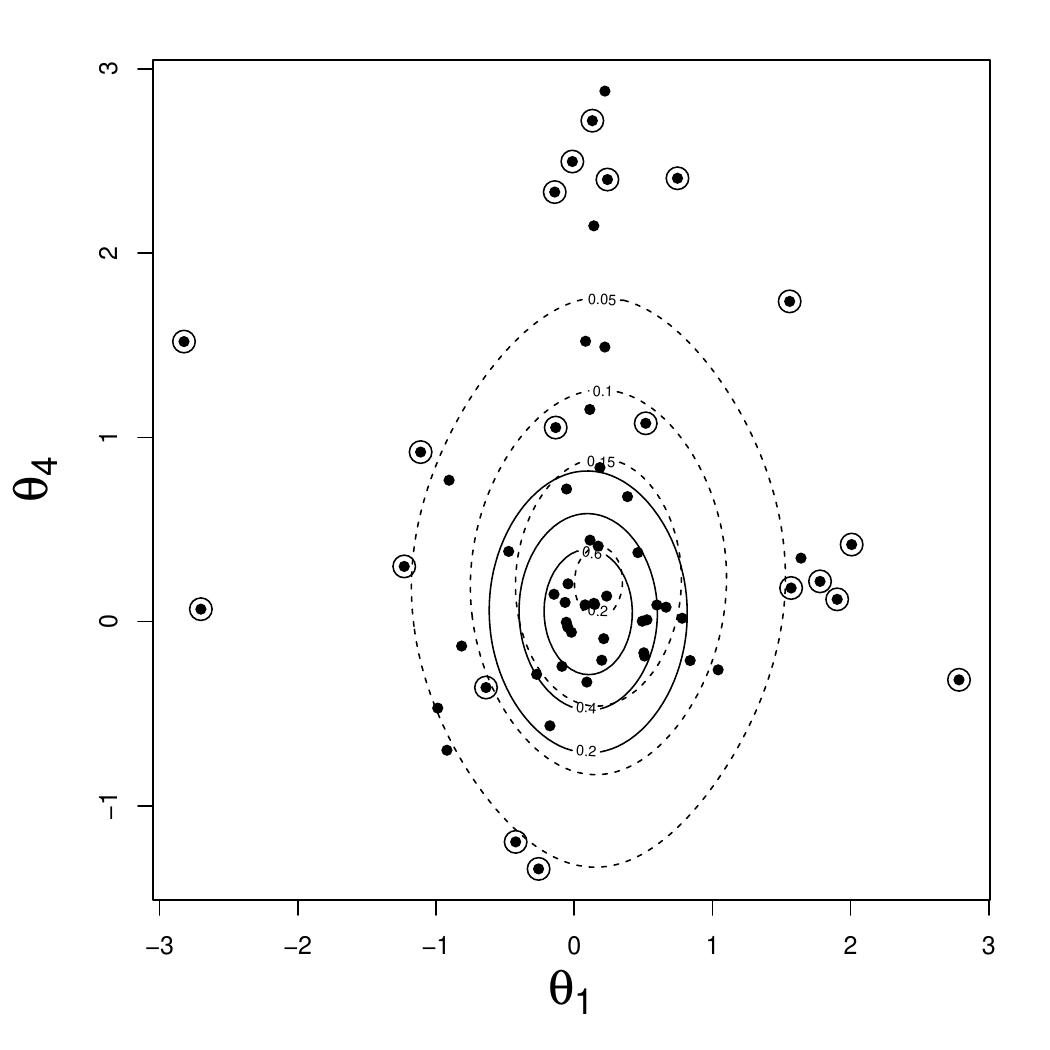}\\
	\includegraphics[width=0.31\textwidth]{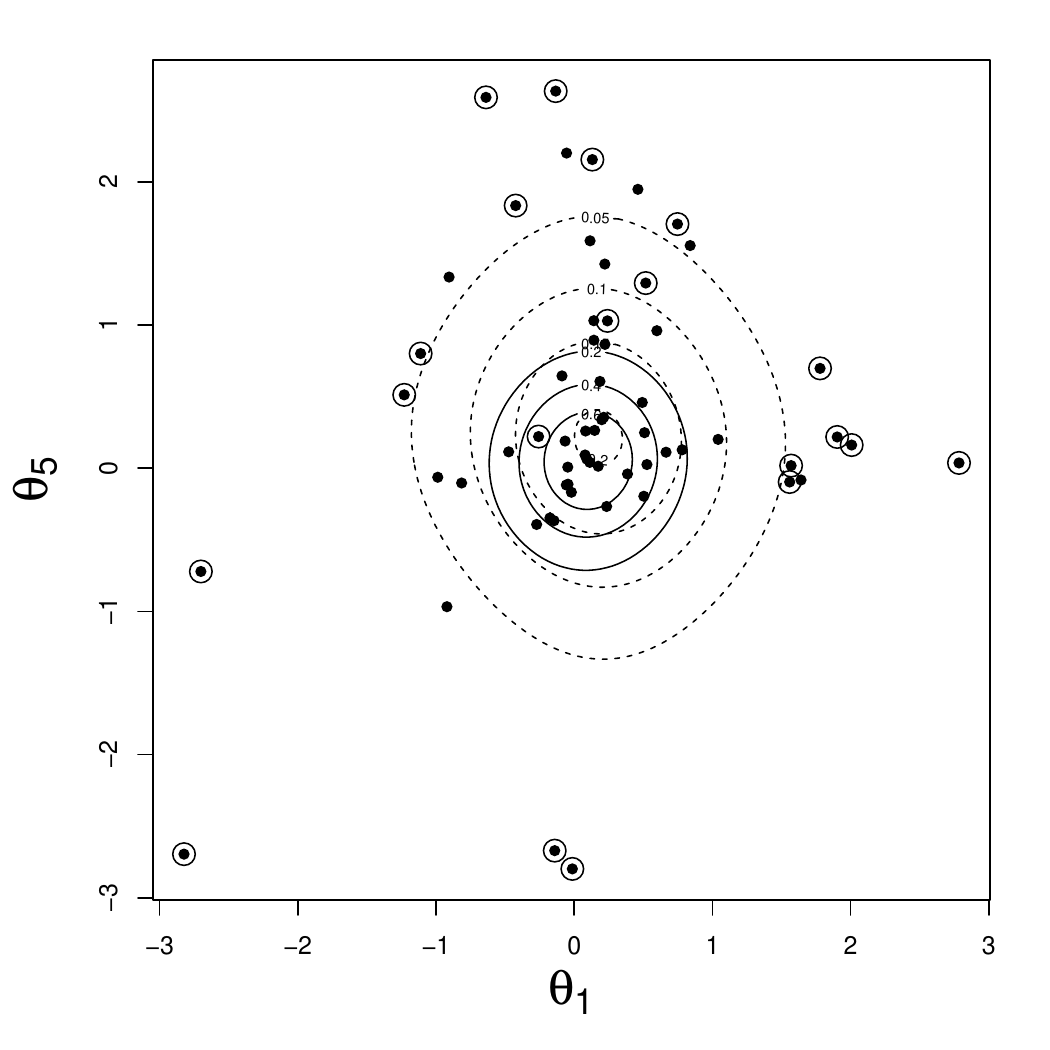}
	\includegraphics[width=0.31\textwidth]{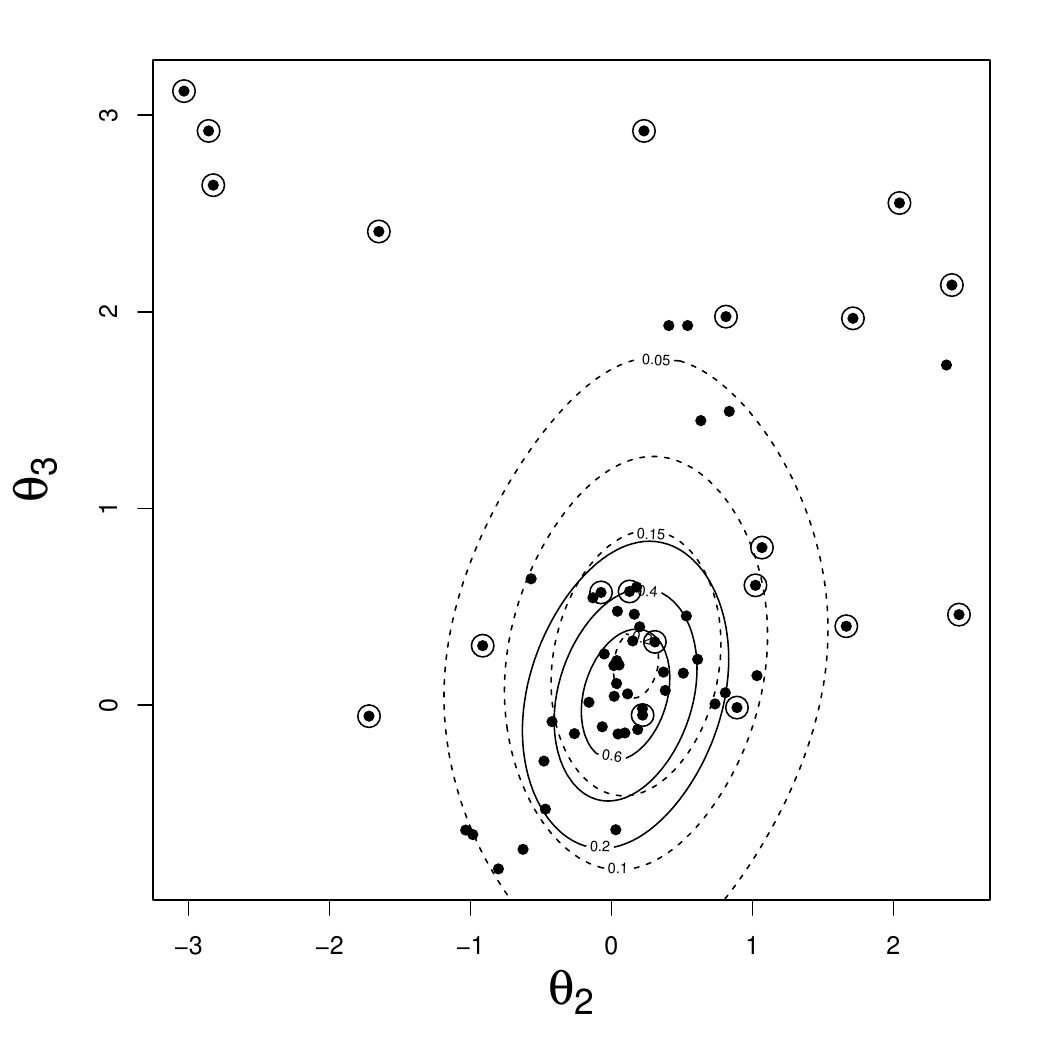}
	\includegraphics[width=0.31\textwidth]{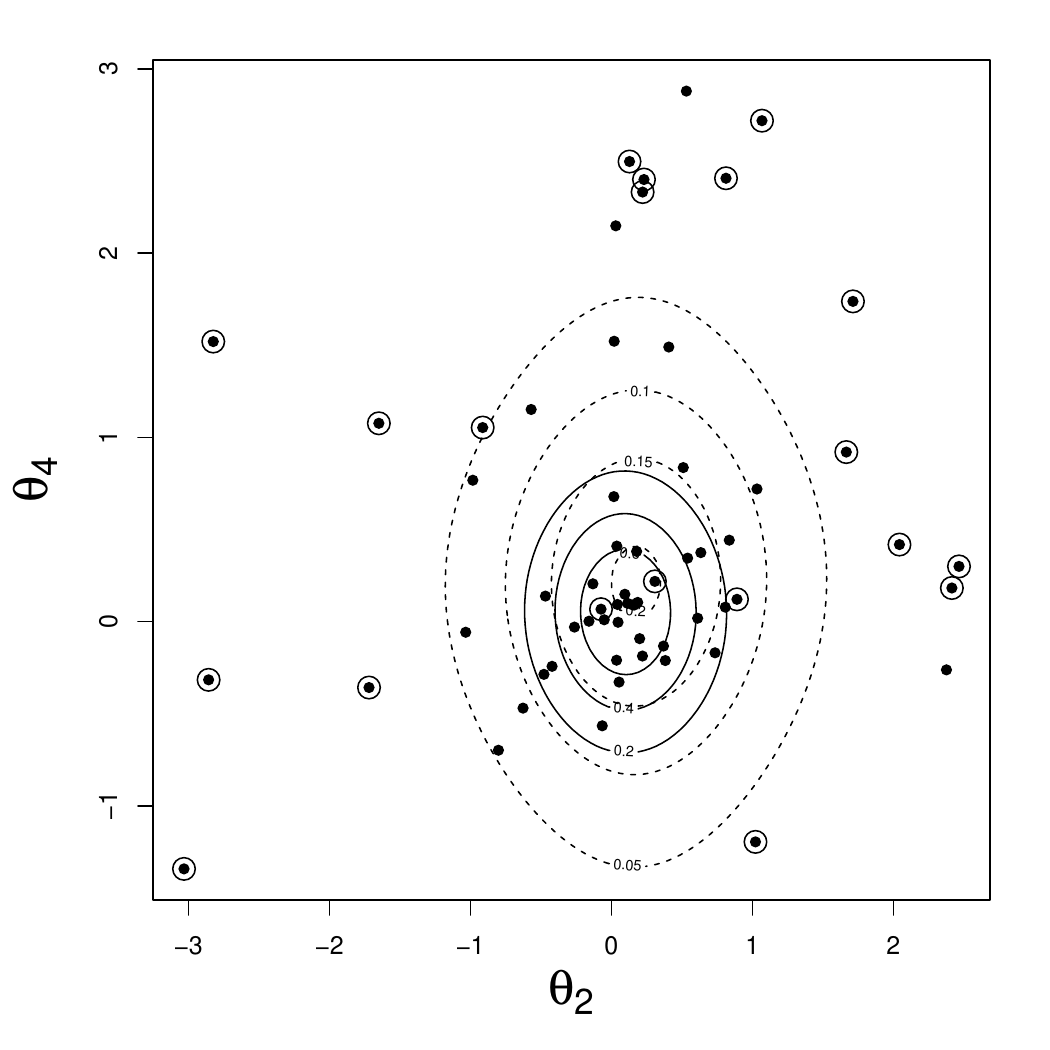}\\
	\includegraphics[width=0.31\textwidth]{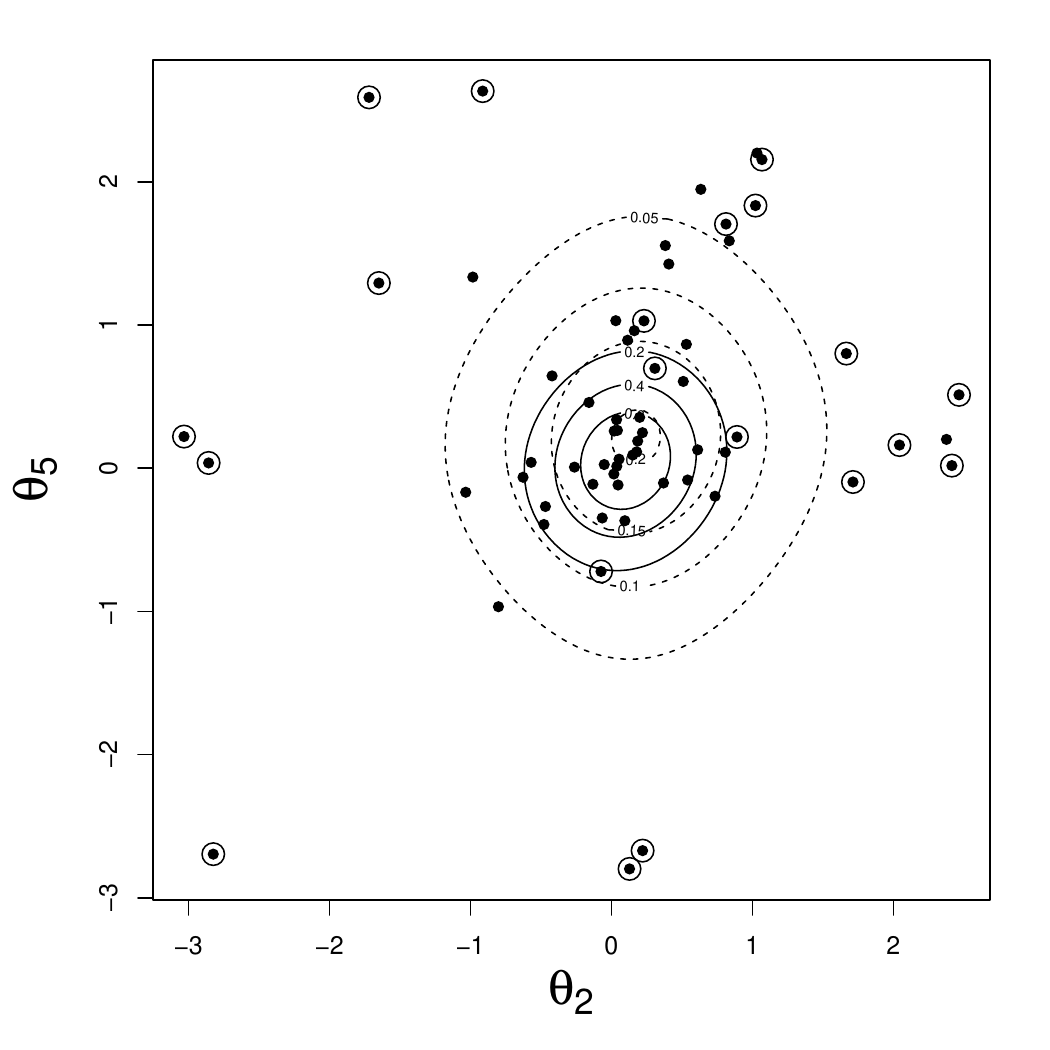}
	\includegraphics[width=0.31\textwidth]{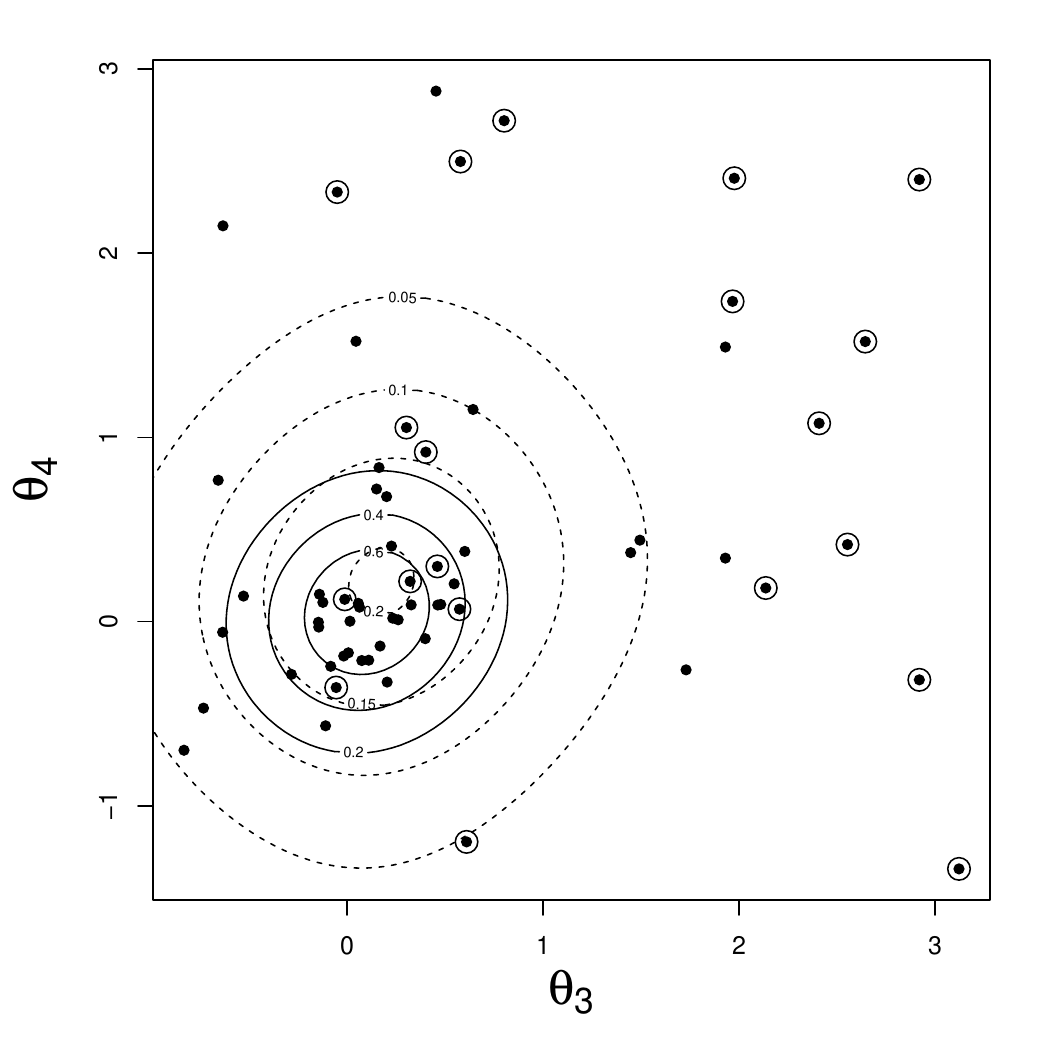}
	\includegraphics[width=0.31\textwidth]{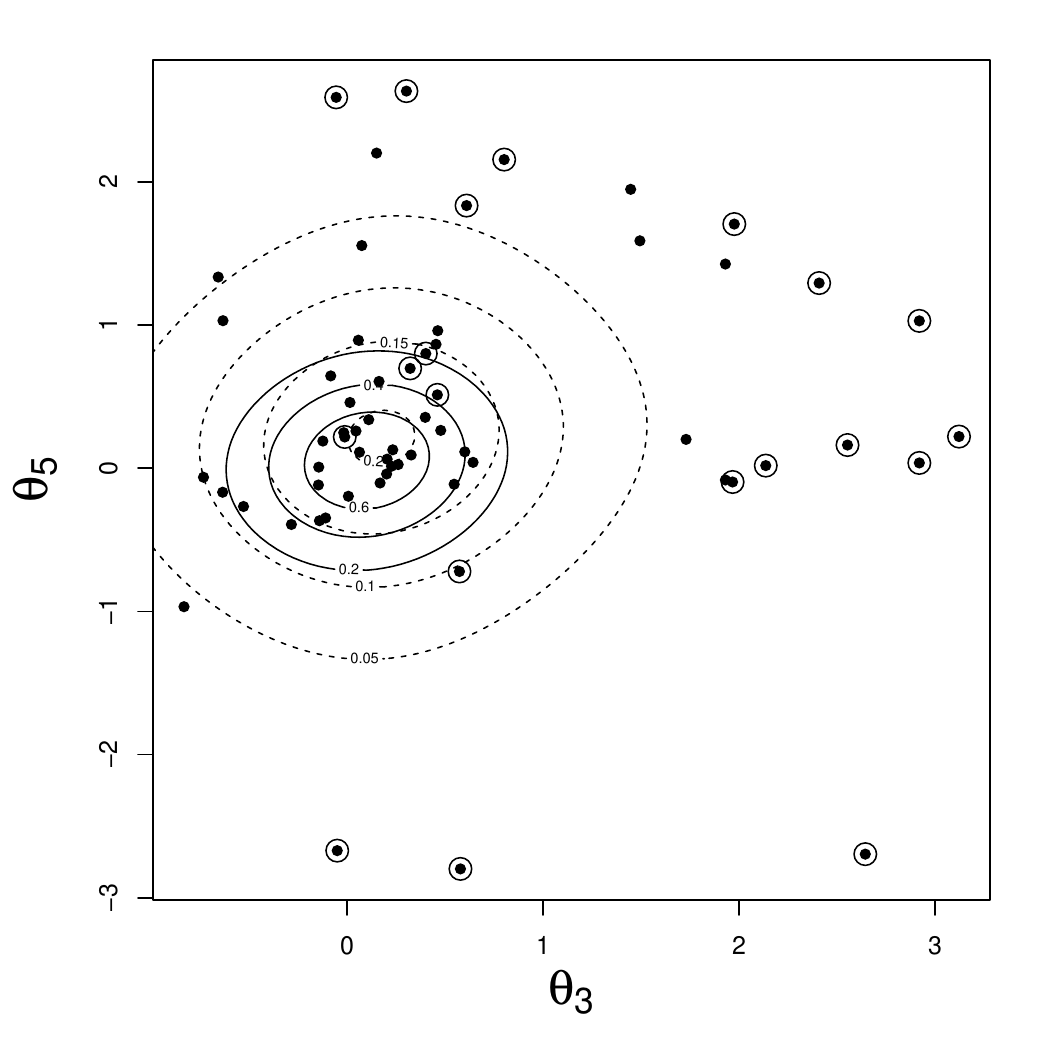}\\
	\includegraphics[width=0.31\textwidth]{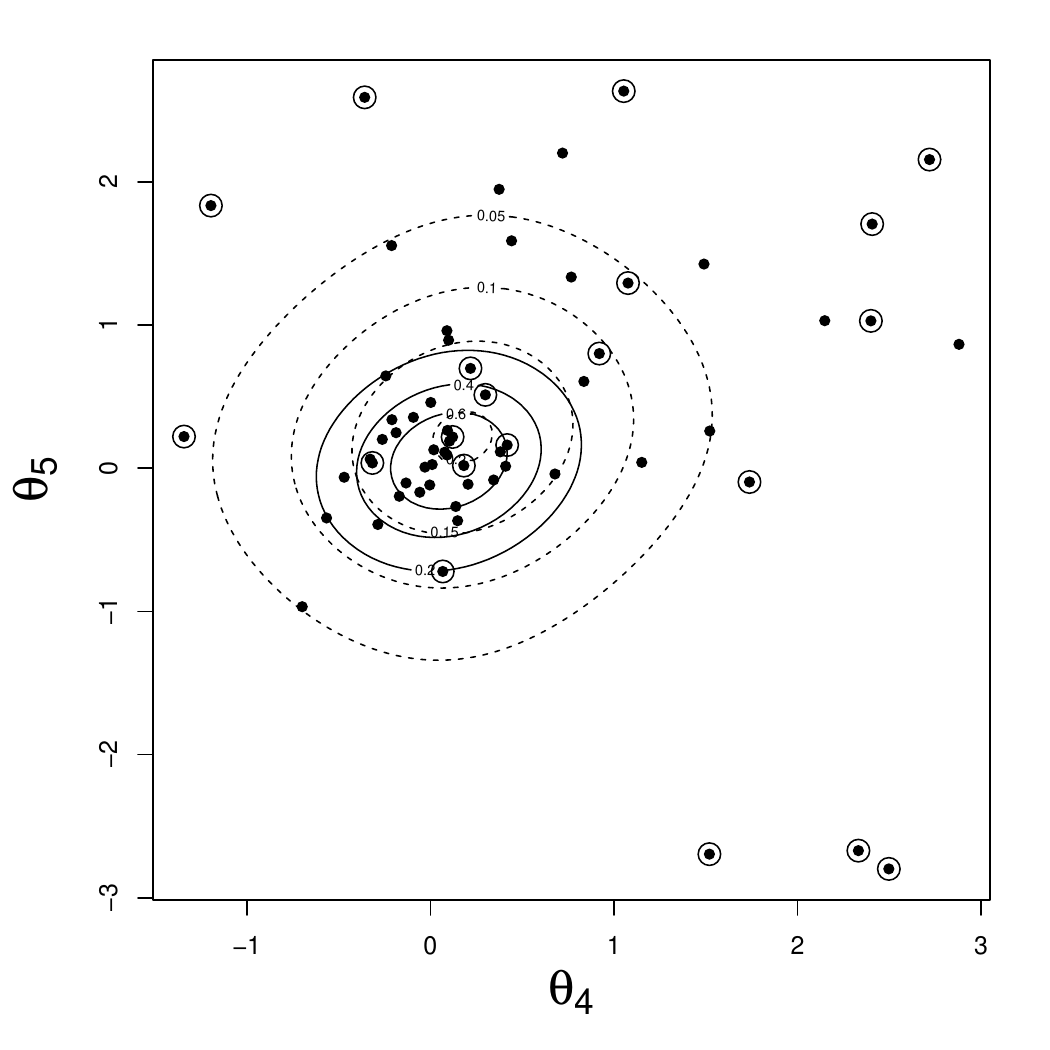}
	\includegraphics[width=0.31\textwidth]{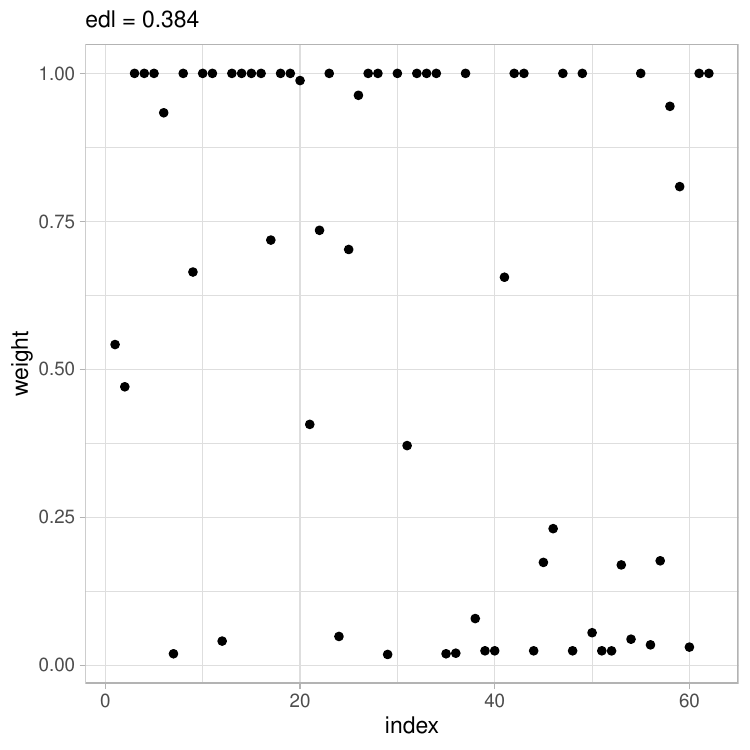}
	\includegraphics[width=0.31\textwidth]{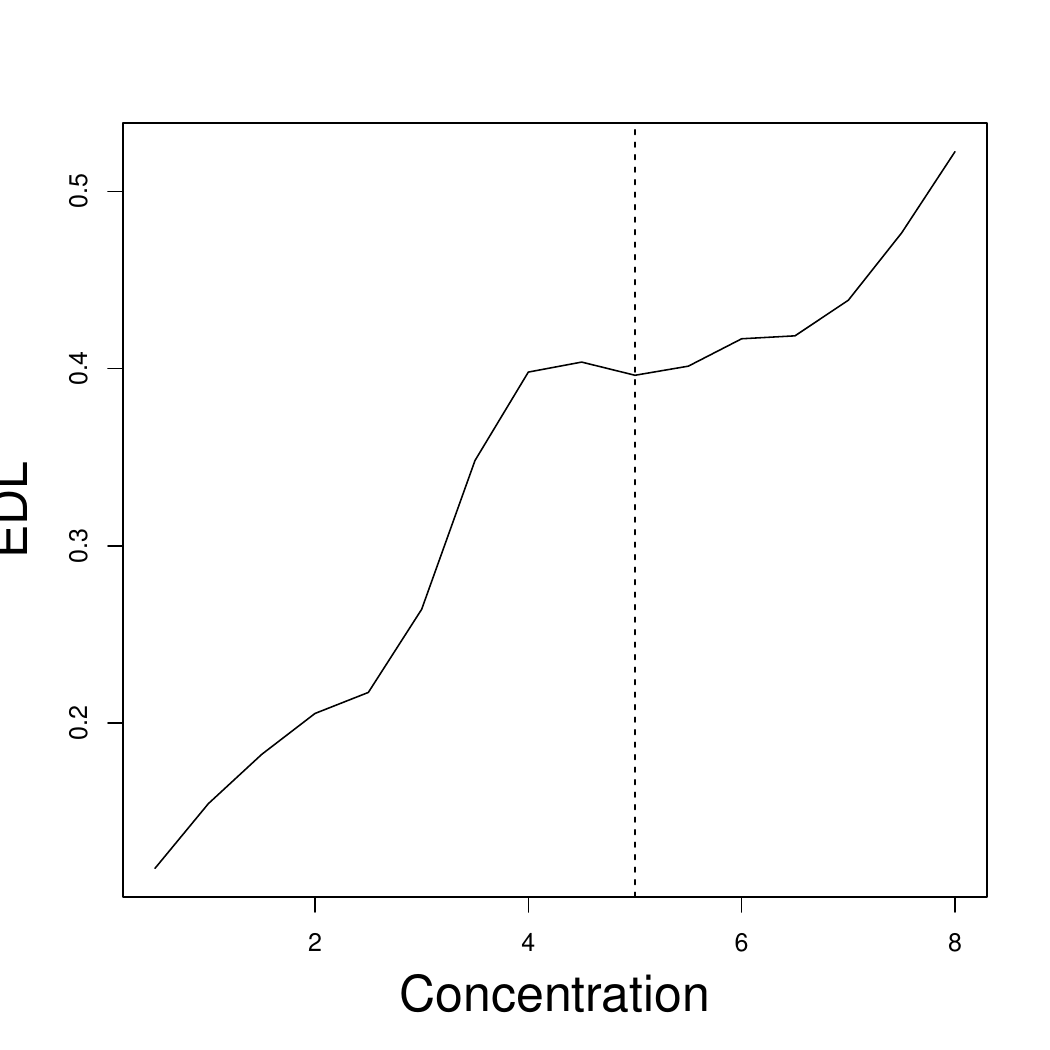}
	\caption{Wind data. Paired scatterplots with marginal bivariate fitted contour densities superimposed: WLE (solid line), MLE (dashed line). The middle panel in the bottom row gives the weights. The right panel in the bottom row gives the monitoring of the empirical downweighting level. The vertical line gives the selected bandwidth.}
	\label{fig:7}
\end{figure}

The data consist of wind direction measurements recorded at a meteorological station (at ``Col De La Roa'') in the Italian Alps. Observations were taken every 15 minutes over 62 consecutive days, from January 29, 2001 to March 31, 2001, between 3:00 a.m. and 4:00 a.m. inclusive. 
The dataset is available in the {\tt R} package {\tt circular} \citep{AgostinelliLund2017}. 
Our aim is to study the associations among measurements at the five different time points; consequently, we analyze 62 daily measurement of a five-dimensional circular random vector.

Figure~\ref{fig:7} shows the bivariate marginal fits obtained via WLE (based on a GKL RAF with $k^*=5$) and MLE, superimposed on the paired data. The corresponding fitted parameter values are reported in Table~\ref{tab:2}. 
Overall, the robust fitted model exhibits larger concentration parameters across all dimensions. Differences are also observed in the dependence structure, as summarized by the $\lambda_{ij}$ parameters.
There are several measurements that show an anomalous behavior in some dimensions. These data inadequacies are properly downweighted by weighted likelihood resulting in differently shaped bivariate marginal contours: points whose weight is not greater than 0.2 have been highlighted.
Weights are given in middle panel in the bottom row. The right panel in the bottom row displays the monitoring plot of the empirical downweighting level as the concentration bandwidth varies. Moving to the right, the fit becomes more robust, as the concentration bandwidth increases, and the empirical downweighting level grows, accordingly. The concentration bandwidth is selected where the empirical downweighting level first stabilizes.

\begin{table}[ht]
\centering
\begin{tabular}{rrrrrrrrrrr}
  \hline
 & \multicolumn{5}{c|}{$\vect{\mu}$} & \multicolumn{5}{c}{$\vect{\kappa}$}\\
  \hline
MLE & 0.175 & 0.214 & 0.37  & 0.299  & 0.413  & 1.805 & 1.449  & 1.536 & 1.584 & 1.473 \\
WLE & 0.101 & 0.053 & 0.117 & 0.169  & 0.255  & 5.631 & 4.969  & 4.543 & 3.184 & 3.424 \\
  \hline
 & $\lambda_{12}$ & $\lambda_{13}$ &  $\lambda_{14}$ & 
$\lambda_{15}$ & $\lambda_{23}$ & $\lambda_{24}$ &  $\lambda_{25}$ &  $\lambda_{34}$ &  $\lambda_{35}$ & $\lambda_{45}$ \\ \hline
MLE & 0.211 & 0.286 & 0.254 & 0.041  & 0.027  & 0.187 & -0.062 & 0.079 & 0.134 & 0.236 \\
WLE & 1.456 & 0.9   & 1.373 & -0.061 & -0.062 & 0.488 & 0.138  & 0.498 & 0.476 & 0.839 \\
   \hline
\end{tabular}
\caption{Wind data. Parameter estimates from WLE and MLE.}
\label{tab:2}
\end{table}

\section{Conclusions}
\label{sec:7}

This paper addresses the problem of robust parameter estimation for a multivariate von Mises sine distribution, a task of practical importance given the widespread use of von Mises models in the analysis of multivariate circular data and the well-known sensitivity of maximum likelihood estimation to contamination.  

We propose a robust fitting procedure based on weighted likelihood estimation, where downweighting is driven by a probabilistic notion of outlyingness. This approach is particularly well suited to circular data, as it avoids the introduction of ad hoc geometric distances on the torus, which are often difficult to justify and interpret statistically \citep[see][for a review]{agostinelli2024weighted}. The resulting estimator retains the main advantages of likelihood-based methods under the model while substantially improving robustness in the presence of outliers.

Our numerical studies demonstrate that the proposed WLE provides stable and reliable inference across a range of contamination scenarios and dimensions, albeit at the cost of some efficiency loss in the estimation of concentration parameters. This trade-off is intrinsic to robust procedures and is consistent with what is observed in related robust likelihood frameworks.

In the high-concentration regime of the von Mises sine model, two classical approximations are commonly employed: the normal approximation and the concentrated multivariate sine model. Although the latter is often regarded as a more accurate approximation to the true von Mises sine density and is widely used in applications, our simulation results indicate that, for inferential purposes, treating the data as arising from a Wrapped Normal distribution can lead to improved performance. In particular, the Wrapped Normal--based WLE yields more accurate estimation of dependence and concentration parameters and more reliable inferential procedures under contamination.
Furthermore, the Wrapped Normal (WN) model provided a more effective framework for balancing the trade-off between efficiency and robustness, as it allowed for a more precise calibration of the empirical downweighting level.

These findings suggest that, in robust inference for highly concentrated multivariate circular data, the analytical tractability and favorable inferential properties of the Wrapped Normal model may outweigh the benefits of a more faithful---but analytically complex---approximation of the von Mises sine distribution. This observation opens interesting directions for further research on robustness--efficiency trade-offs in circular and directional statistics.

\section*{Declarations}

Luca Greco has been partially funded by the PRIN funding scheme of the
Italian Ministry of University and Research (Grant no. 2022LANNKC CUP
E53D23005810006)

\noindent Claudio Agostinelli has been partially funded by the PRIN funding scheme
of the Italian Ministry of University and Research (Grant no. 2022FZY9PM
CUP C53C24000740006)

\section{Appendix}\label{secA1}

\begin{figure}[h]
	\centering
	\includegraphics[width=0.9\textwidth]{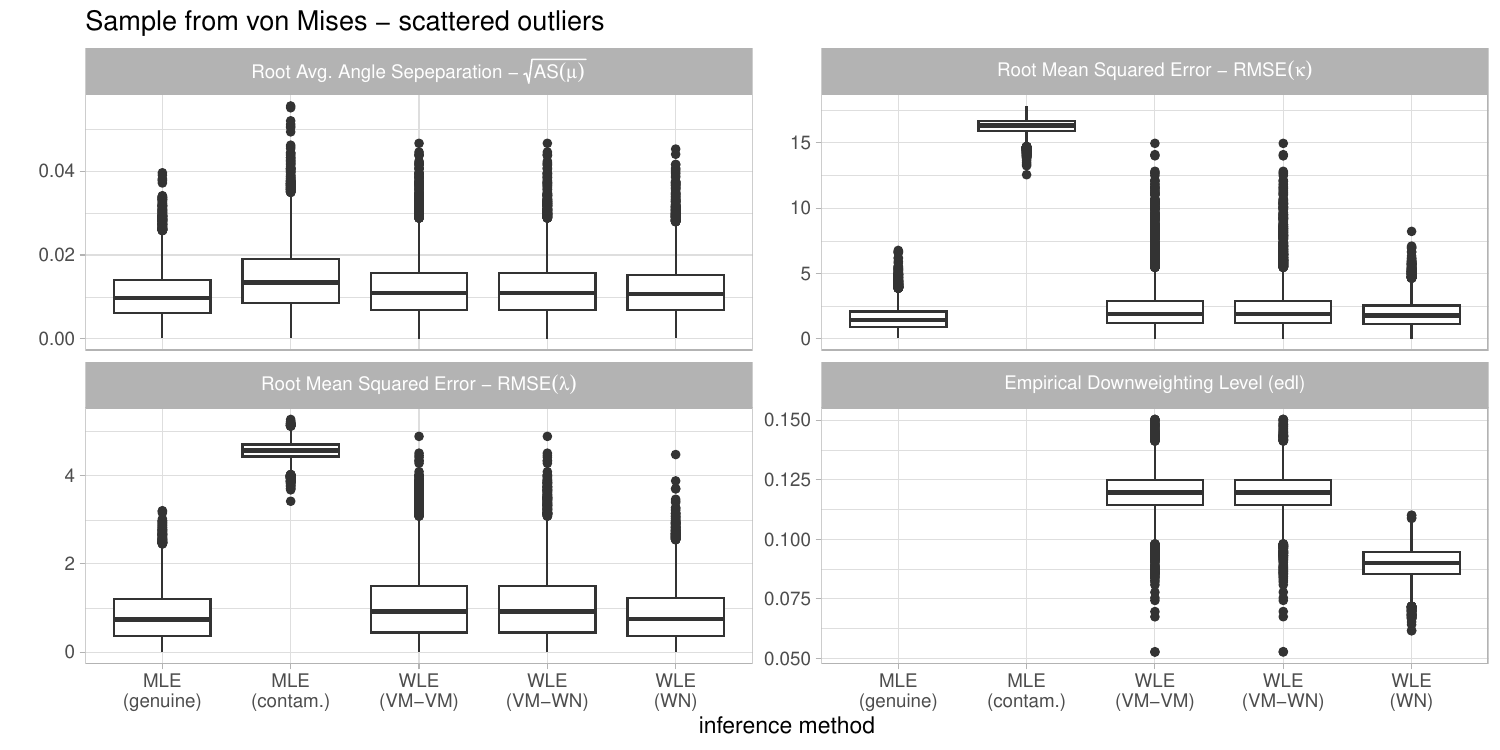}
	\caption{Monte Carlo simulation (bivariate). Empirical distributions of the estimation errors for parameters $\vect{\mu}$, $\vect{\kappa}$, and $\lambda$, together with the empirical downweighting level (edl). Different estimation methods are displayed along the $x-$axis. Results are based on samples of size $n=300$ from a bivariate von Mises sine distribution with scattered outliers.}
	\label{fig:sim2d-acc-vm-sc}
\end{figure}

\begin{figure}[h]
	\centering
	\includegraphics[width=0.9\textwidth]{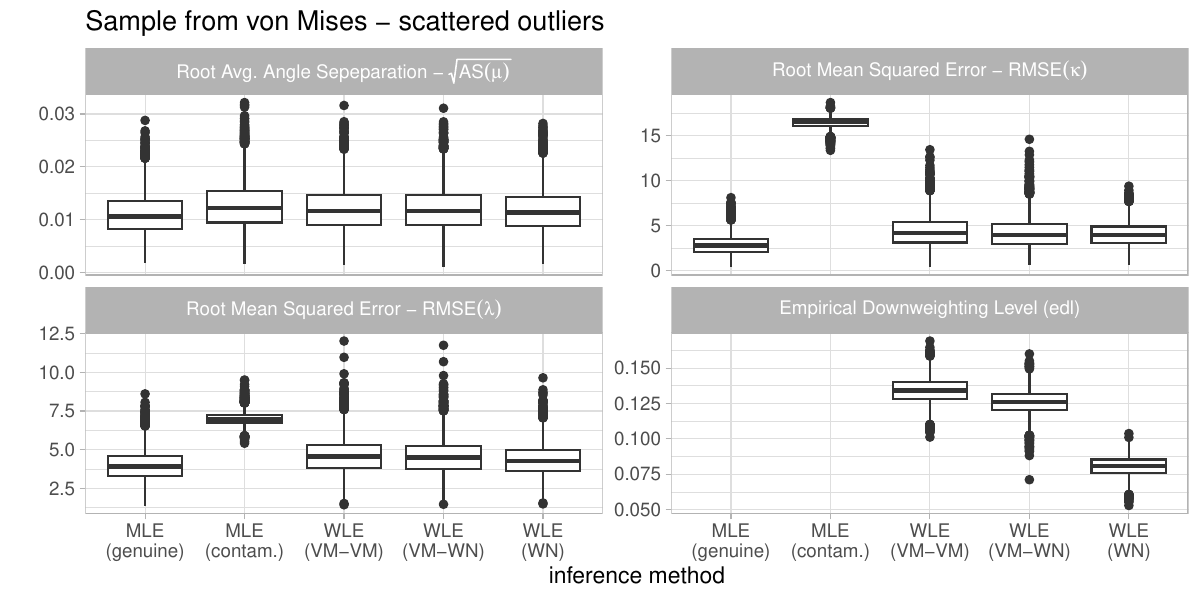}
	\caption{Monte Carlo simulation (five-dimensional). Empirical distributions of the estimation errors for parameters $\vect{\mu}$, $\vect{\kappa}$, and $\lambda$, together with the empirical downweighting level (edl). Different estimation methods are displayed along the $x-$axis. Results are based on samples of size $n=300$ from a five-dimensional von Mises distribution with scattered outliers.}
	\label{fig:sim5d-acc-vm-sc}
\end{figure}

\begin{figure}[h]
	\centering
	\includegraphics[width=0.9\textwidth]{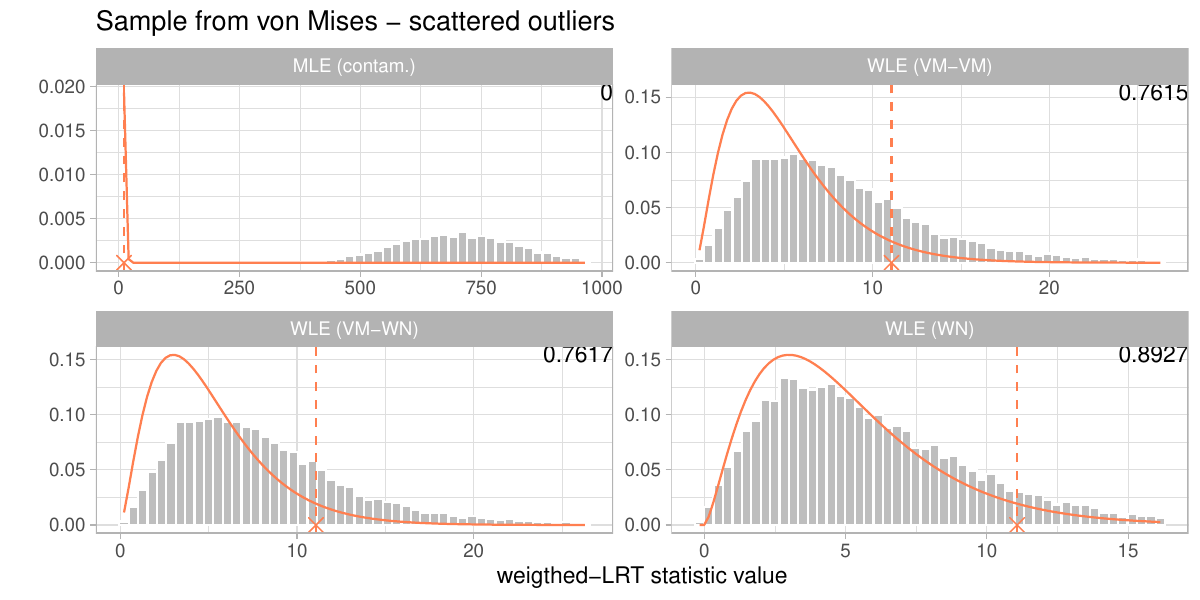}
	\caption{Monte Carlo simulations (bivariate). Sampling distributions for the WLRT (LRT for MLE under contamination) statistic and the $\chi^2-$ density superimosed. Empirical coverage at level 0.95 printed on the top-right of each subplot. Results are based on samples of size $n=300$ from a bivariate von Mises sine distribution with scattered outliers.}
	\label{fig:wlrt-sc}
\end{figure}

\begin{figure}[h]
	\centering
	\includegraphics[width=0.9\textwidth]{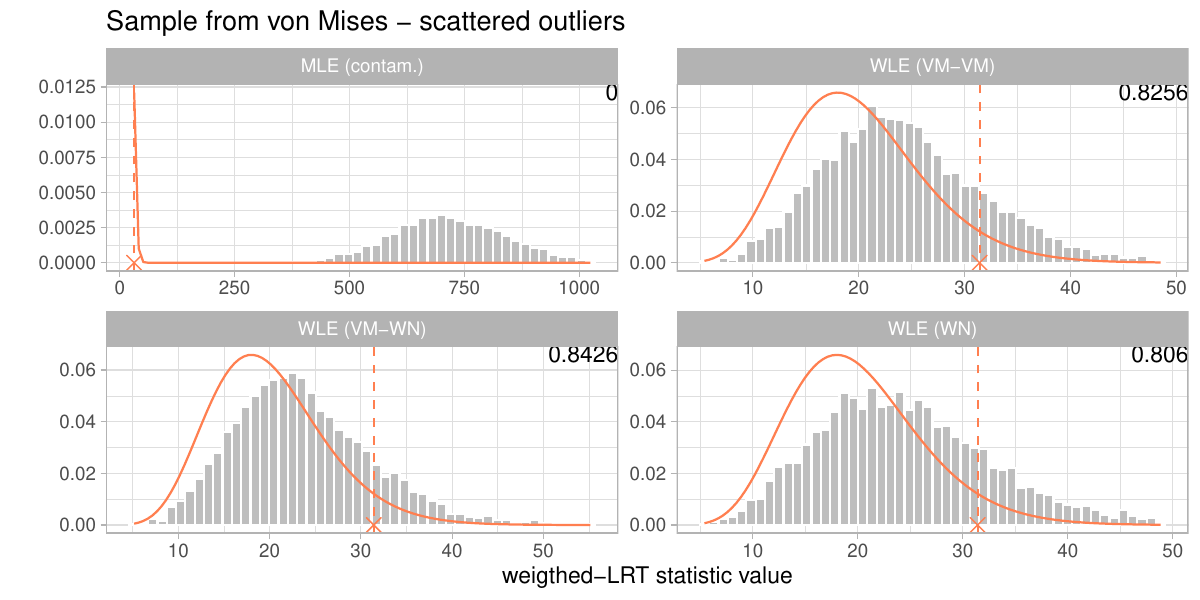}
	\caption{Monte Carlo simulations (five-dimensional). Sampling distributions for the WLRT (LRT for MLE under contamination) statistic and the $\chi^2-$ density superimosed. Empirical coverage at level 0.95 printed on the top-right of each subplot. Results are based on samples of size $n=300$ from a five-dimensional von Mises sine distribution with scattered outliers.}
	\label{fig:wlrt-5d-sc}
\end{figure}

In this appendix we collect additional figures and tables from the Monte Carlo simulations described in Sec.~\ref{sec:5}. 
Tables~\ref{tab:coverage-vm} and \ref{tab:coverage-vm-5}, in particular, summarize the coverage for the (W)LRT statistics, respectively, for the bivariate and five-dimensional simulations.
Figures~\ref{fig:sim2d-acc-vm-sc}-\ref{fig:sim5d-acc-vm-sc} show the empirical distributions for the different accuracy measures in the scattered-outliers contamination scenario, again for $p=2$ and $p=5$.
For the same outliers type, we show also the sampling distributions for the different (W)LRT statistics in Figures~\ref{fig:wlrt-sc} and \ref{fig:wlrt-5d-sc}.
Note when outliers are scattered, and in particular in the five-dimensional scenario, the performance of all the three WLE approaches are very similar.

\begin{figure}[h]
	\centering
	\includegraphics[width=0.9\textwidth]{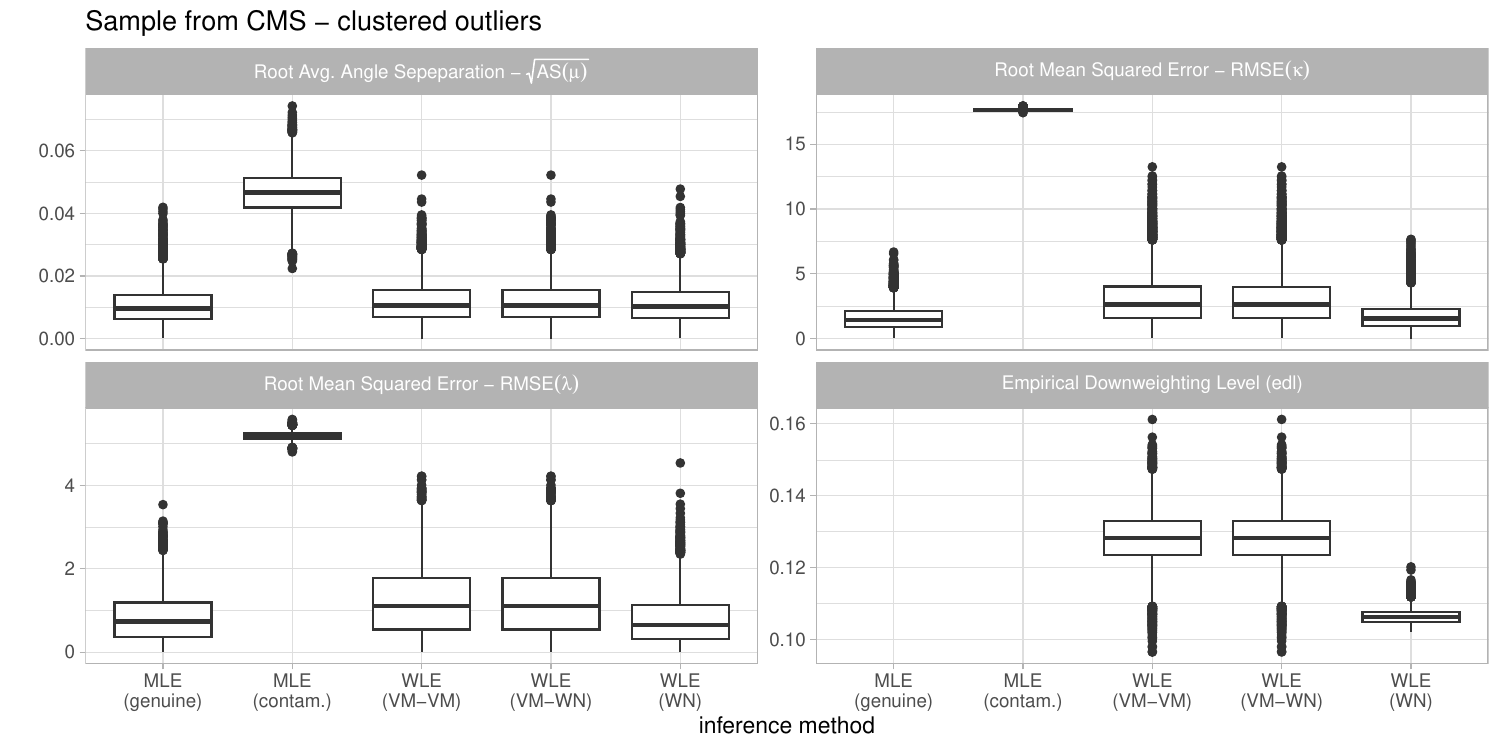}
	\caption{Monte Carlo simulation (bivariate). Empirical distributions of the estimation errors for parameters $\vect{\mu}$, $\vect{\kappa}$, and $\lambda$, together with the empirical downweighting level (edl). Different estimation methods are displayed along the $x-$axis. Results are based on samples of size $n=300$ from a bivariate concentrated multivariate sine (CMS) distribution with clustered outliers.}
	\label{fig:sim2d-acc-cms-gr}
\end{figure}

\begin{figure}[h]
	\centering
	\includegraphics[width=0.9\textwidth]{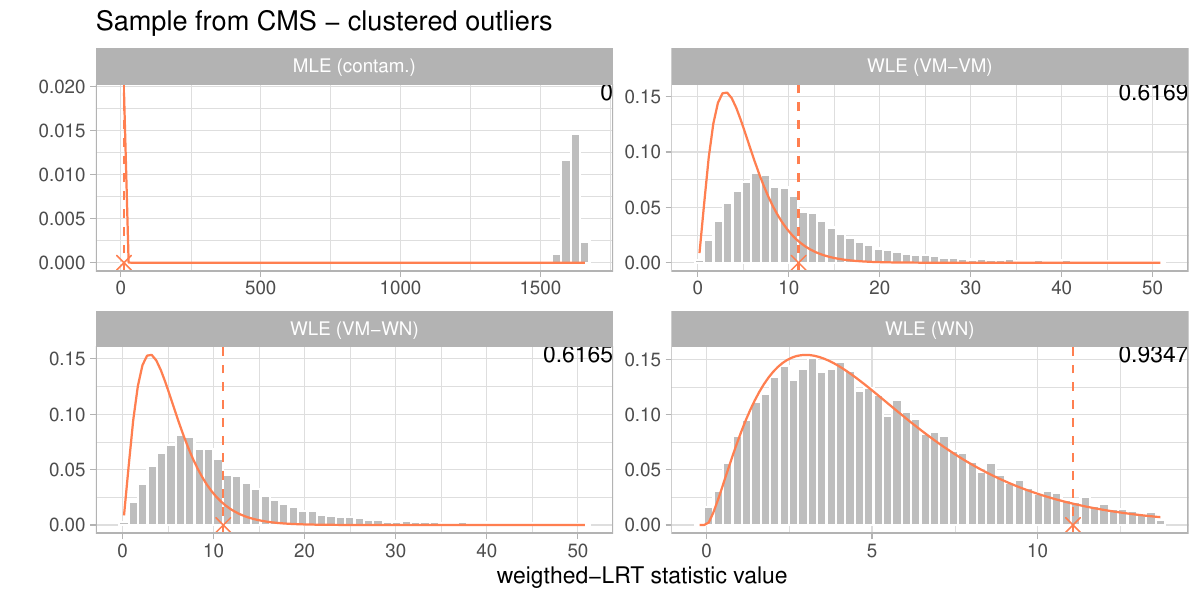}
	\caption{Monte Carlo simulations (bivariate). Sampling distributions for the WLRT (LRT for MLE under contamination) statistic and the $\chi^2-$ density superimosed. Empirical coverage at level 0.95 printed on the top-right of each subplot. Results are based on samples of size $n=300$ from a bivariate concentrated multivariate sine (CMS) distribution with clustered outliers.}
	\label{fig:wlrt-cms-gr}
\end{figure}

\begin{figure}[h]
	\centering
	\includegraphics[width=0.9\textwidth]{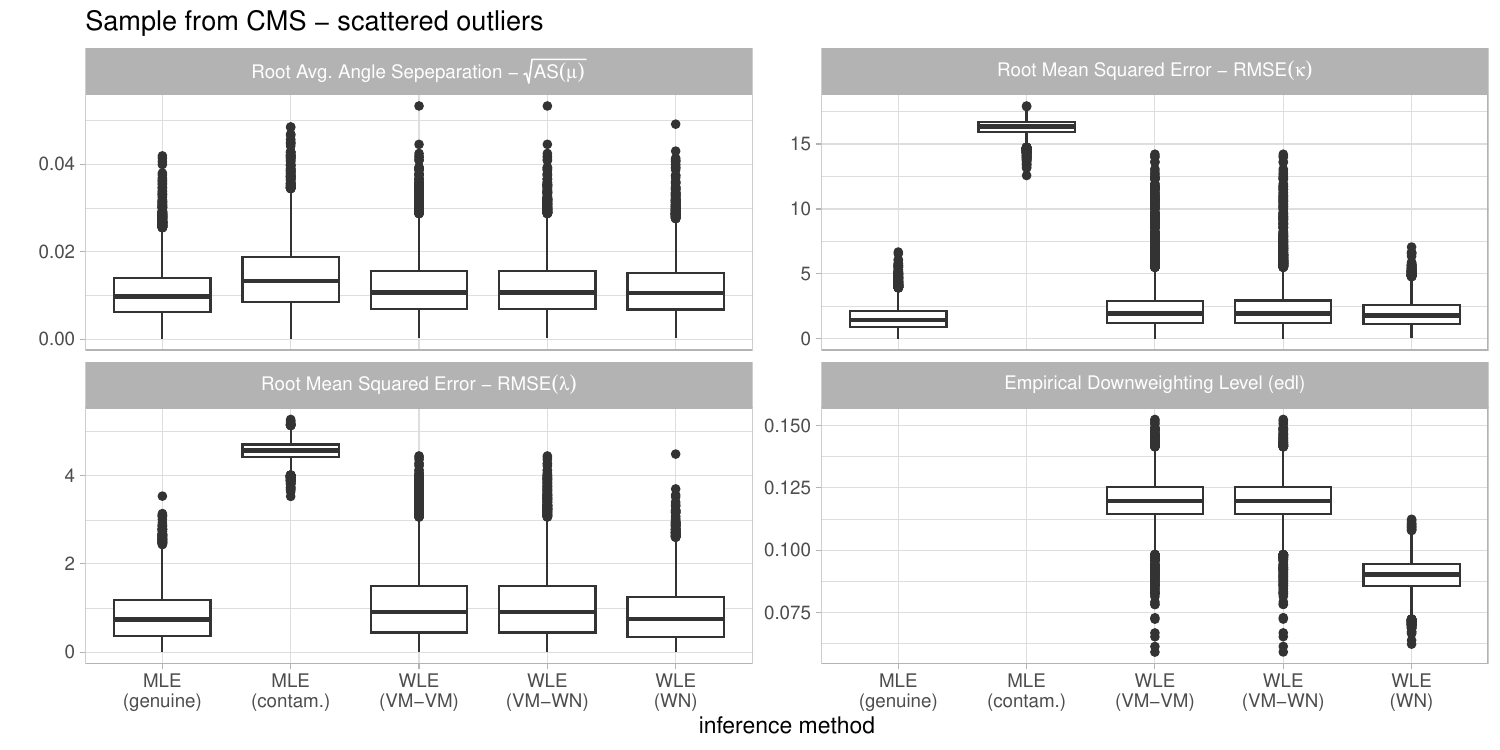}
	\caption{Monte Carlo simulation (bivariate). Empirical distributions of the estimation errors for parameters $\vect{\mu}$, $\vect{\kappa}$, and $\lambda$, together with the empirical downweighting level (edl). Different estimation methods are displayed along the $x-$axis. Results are based on samples of size $n=300$ from a bivariate concentrated multivariate sine (CMS) distribution with scattered outliers.}
	\label{fig:sim2d-acc-cms-sc}
\end{figure}

\begin{figure}[h]
	\centering
	\includegraphics[width=0.9\textwidth]{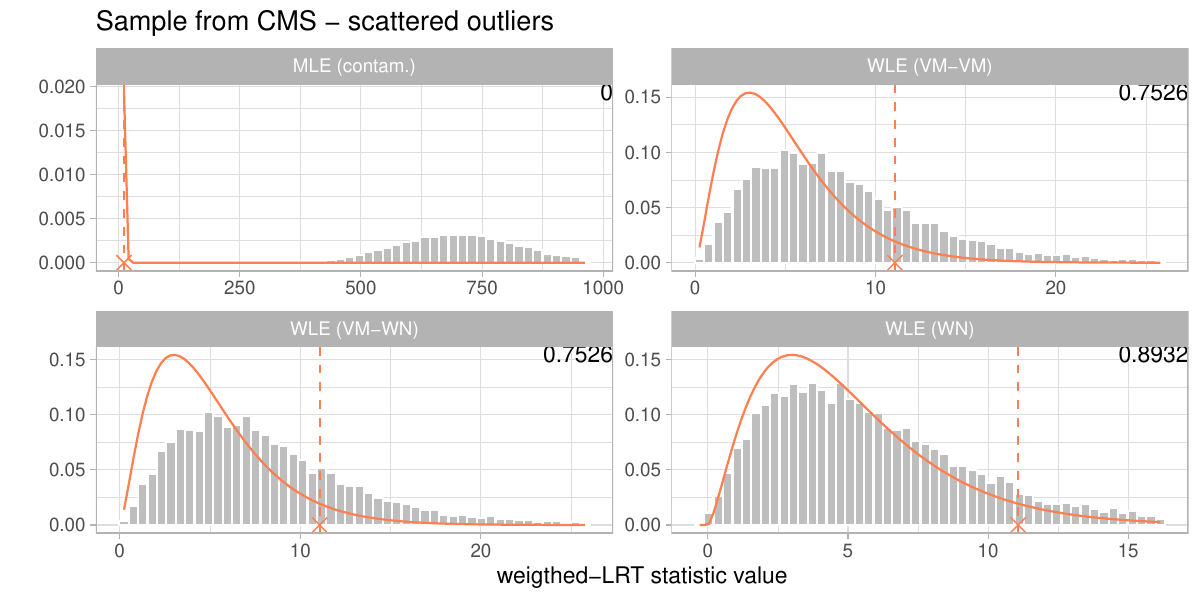}
	\caption{Monte Carlo simulations (bivariate). Sampling distributions for the WLRT (LRT for MLE under contamination) statistic and the $\chi^2-$ density superimosed. Empirical coverage at level 0.95 printed on the top-right of each subplot. Results are based on samples of size $n=300$ from a bivariate concentrated multivariate sine (CMS) distribution with scattered outliers.}
	\label{fig:wlrt-cms-sc}
\end{figure}

\begin{figure}[h]
	\centering
	\includegraphics[width=0.9\textwidth]{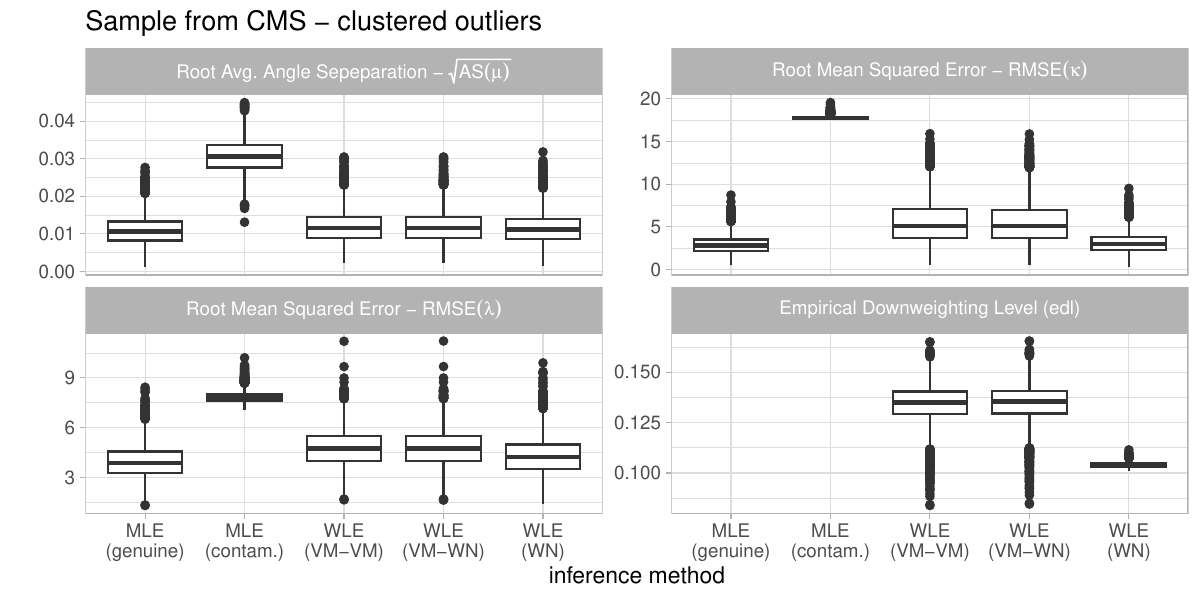}
	\caption{Monte Carlo simulation (five-dimensional). Empirical distributions of the estimation errors for parameters $\vect{\mu}$, $\vect{\kappa}$, and $\lambda$, together with the empirical downweighting level (edl). Different estimation methods are displayed along the $x-$axis. Results are based on samples of size $n=300$ from a five-dimensional concentrated multivariate sine (CMS) distribution with clustered outliers.}
	\label{fig:sim5d-acc-cms-gr}
\end{figure}

\begin{figure}[h]
	\centering
	\includegraphics[width=0.9\textwidth]{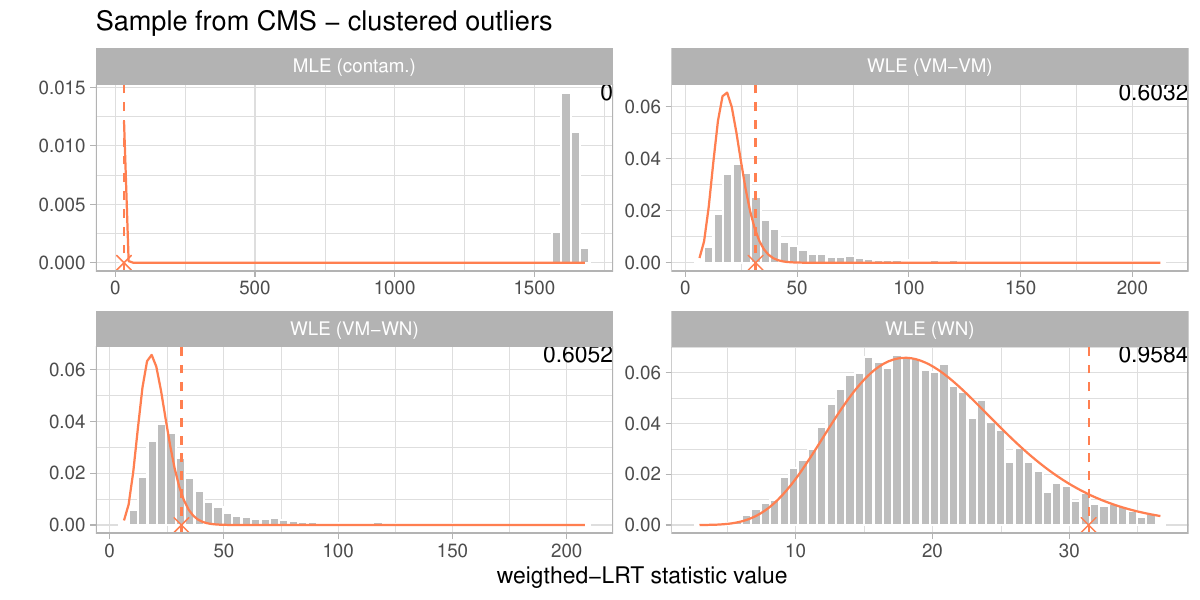}
	\caption{Monte Carlo simulations (five-dimensional). Sampling distributions for the WLRT (LRT for MLE under contamination) statistic and the $\chi^2-$ density superimosed. Empirical coverage at level 0.95 printed on the top-right of each subplot. Results are based on samples of size $n=300$ from a five-dimensional concentrated multivariate sine (CMS) distribution with clustered outliers.}
	\label{fig:wlrt-5d-cms-gr}
\end{figure}

\begin{figure}[h]
	\centering
	\includegraphics[width=0.9\textwidth]{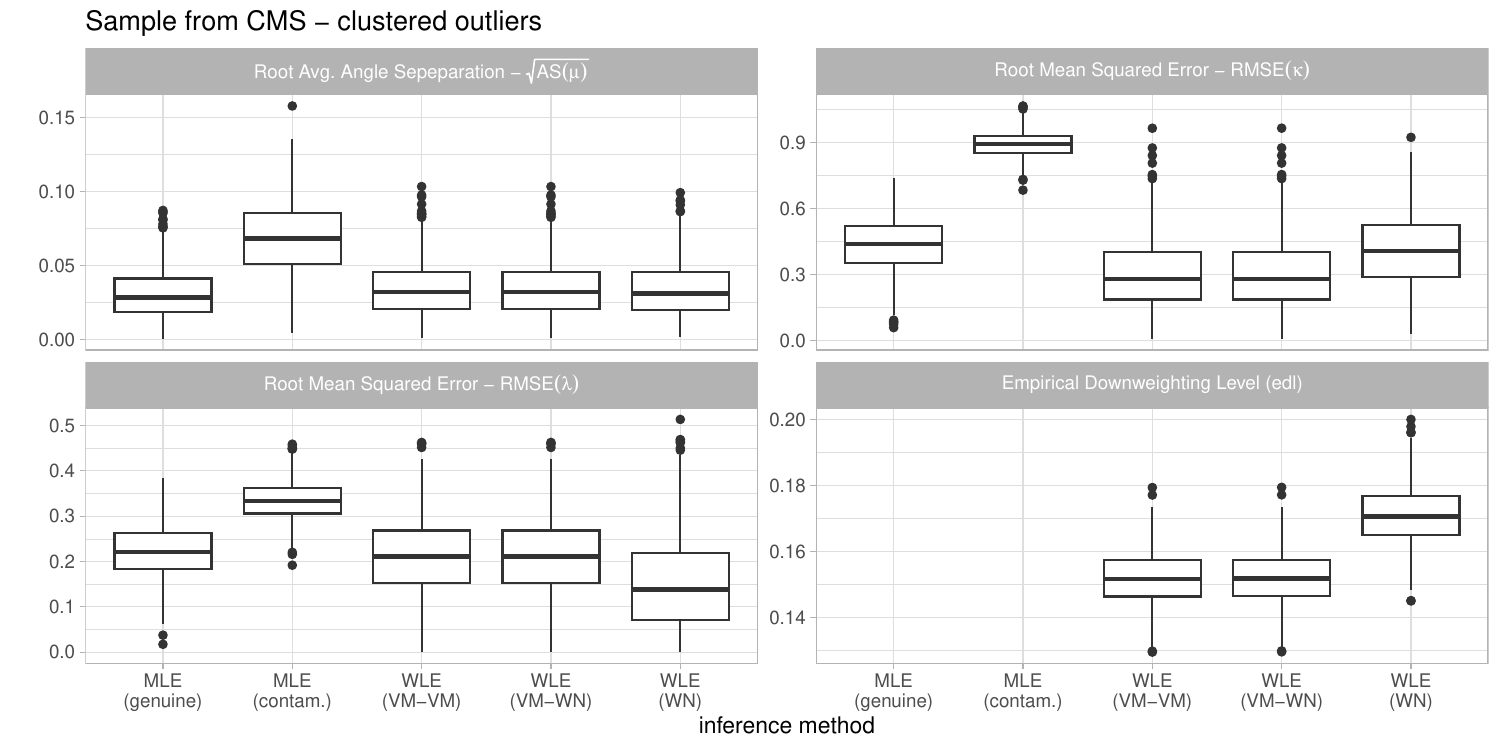}
	\caption{Monte Carlo simulation (bivariate). Empirical distributions of the estimation errors for parameters $\vect{\mu}$, $\vect{\kappa}$, and $\lambda$, together with the empirical downweighting level (edl). Different estimation methods are displayed along the $x-$axis. Results are based on samples of size $n=300$ from a bivariate CMS distribution with $\vect{\kappa}=(2, 2)$ and $\lambda=0.35$, with scattered outliers.}
	\label{fig:sim2d-bis-gr}
\end{figure}

Finally, we include the results for a third stream of simulations, where the genuine circular data are sampled directly from the concentrated multivariate sine (CMS) distribution of eq.~\eqref{eq:dcms}, instead of the von Mises sine. The random sampling of the CMS is achieved through importance sampling, where the auxiliary distribution if the multivariate central $t-$distribution with the same covariance structure $\Sigma$ of the CMS and $20$ degrees of freedom. 
In the first set of simulations we keep the same parameters, i.e. $\vect{\kappa}=(10, 20)^\top$ and $\lambda=5$ in the bivariate case, and similarly in five-dimension. Figures~\ref{fig:sim2d-acc-cms-gr}--\ref{fig:wlrt-5d-cms-gr} summarize their results.
In the second set of simulations we consider only the two-dimensional case and choose $\vect{\kappa}=(2, 2)^\top$ and $\lambda=0.35$. In this case VM--VM WLE yields more accurate estimates for $\vect{\kappa}$ than WN WLE and, even more accurate than MLE estimates without contamination, see Fig.~\ref{fig:sim2d-bis-gr} and following.

\begin{table}[h]
	\begin{subtable}[h]{0.4\textwidth}
		\centering
		\begin{tabular}{l|l}
			\hline
			& coverage \\ \hline
			MLE (genuine) & 0.9276   \\
			MLE (contam.) & 0        \\
			WLE (VM-VM)   & 0.6155   \\
			WLE (VM-WN)   & 0.6154   \\
			WLE (WN)      & 0.9353  
		\end{tabular}
		\caption{clustered outliers}
	\end{subtable}
	\hfill
	\begin{subtable}[h]{0.4\textwidth}
		\centering
		\begin{tabular}{l|l}
			\hline
			& coverage \\ \hline
			MLE (genuine) & 0.9276   \\
			MLE (contam.) & 0        \\
			WLE (VM-VM)   & 0.7615   \\
			WLE (VM-WN)   & 0.7617   \\
			WLE (WN)      & 0.8927  
		\end{tabular}
		\caption{scattered outliers}
	\end{subtable}
	\caption{Monte Carlo simulations (bivariate). Empirical coverage for the LRT and WLRT for at level 0.95. Results are based on samples of size $n=300$ from a bivariate von Mises sine distribution with contamination.}
	\label{tab:coverage-vm}
\end{table}

\begin{table}[h]
	\begin{subtable}[h]{0.4\textwidth}
		\centering
		\begin{tabular}{l|l}
			\hline
			& coverage \\ \hline
			MLE (genuine) & 0.9496   \\
			MLE (contam.) & 0        \\
			WLE (VM-VM)   & 0.6318   \\
			WLE (VM-WN)   & 0.5982   \\
			WLE (WN)      & 0.9614
		\end{tabular}
		\caption{clustered outliers}
	\end{subtable}
	\hfill
	\begin{subtable}[h]{0.4\textwidth}
		\centering
		\begin{tabular}{l|l}
			\hline
			& coverage \\ \hline
			MLE (genuine) & 0.9496   \\
			MLE (contam.) & 0        \\
			WLE (VM-VM)   & 0.8256   \\
			WLE (VM-WN)   & 0.8426   \\
			WLE (WN)      & 0.806  
		\end{tabular}
		\caption{scattered outliers}
	\end{subtable}
	\caption{Monte Carlo simulations (five-dimensional). Empirical coverage for the LRT and WLRT for at level 0.95. Results are based on samples of size $n=300$ from a five-dimensional von Mises sine distribution with contamination.}
	\label{tab:coverage-vm-5}
\end{table}

% ---
% Figures and tables for samples from CMS
% ---

\begin{table}[h]
	
	\begin{subtable}[h]{0.4\textwidth}
		\centering
		\begin{tabular}{l|l}
			\hline
			& coverage \\ \hline
			MLE (genuine) & 0.929   \\
			MLE (contam.) & 0        \\
			WLE (VM-VM)   & 0.6169   \\
			WLE (VM-WN)   & 0.6165   \\
			WLE (WN)      & 0.9347  
		\end{tabular}
		\caption{clustered outliers}
	\end{subtable}
	\hfill
	\begin{subtable}[h]{0.4\textwidth}
		\centering
		\begin{tabular}{l|l}
			\hline
			& coverage \\ \hline
			MLE (genuine) & 0.929   \\
			MLE (contam.) & 0        \\
			WLE (VM-VM)   & 0.7526   \\
			WLE (VM-WN)   & 0.7526   \\
			WLE (WN)      & 0.8932  
		\end{tabular}
		\caption{scattered outliers}
	\end{subtable}
	\caption{Monte Carlo simulations (bivariate). Empirical coverage for the LRT and WLRT for at level 0.95. Results are based on samples of size $n=300$ from a bivariate concentrated multivariate sine (CMS) distribution with contamination.}
	\label{tab:coverage-cms}
\end{table}

% ---
% 5D
% ---

\begin{figure}[h]
	\centering
	\includegraphics[width=0.9\textwidth]{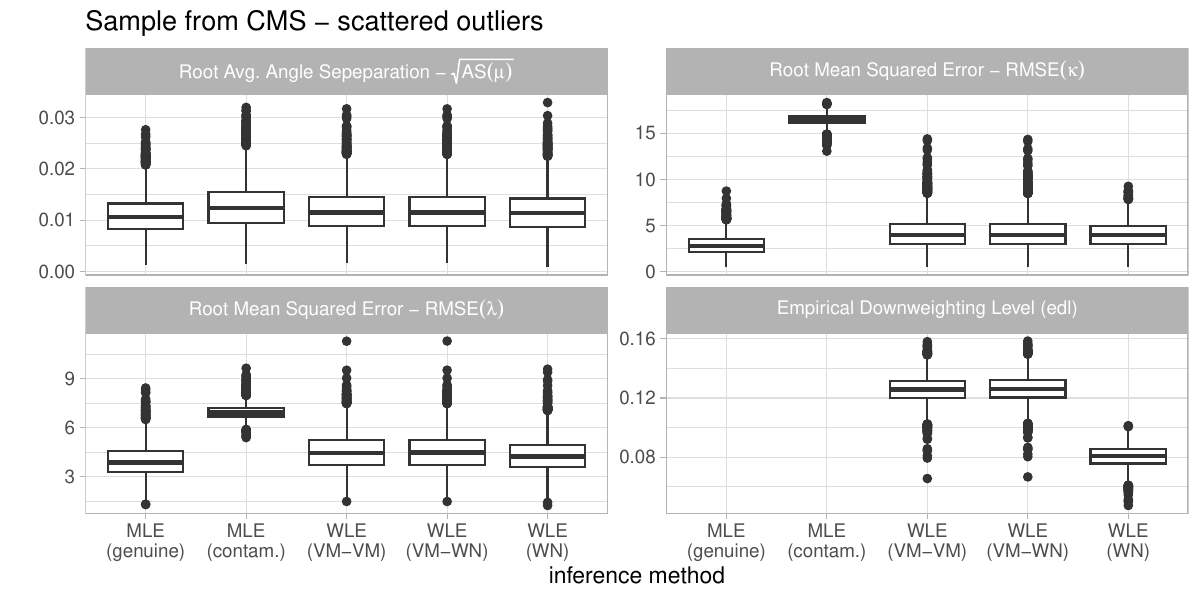}
	\caption{Monte Carlo simulation (five-dimensional). Empirical distributions of the estimation errors for parameters $\vect{\mu}$, $\vect{\kappa}$, and $\lambda$, together with the empirical downweighting level (edl). Different estimation methods are displayed along the $x-$axis. Results are based on samples of size $n=300$ from a five-dimensional concentrated multivariate sine (CMS) distribution with scattered outliers.}
	\label{fig:sim5d-acc-cms-sc}
\end{figure}

\begin{figure}[h]
	\centering
	\includegraphics[width=0.9\textwidth]{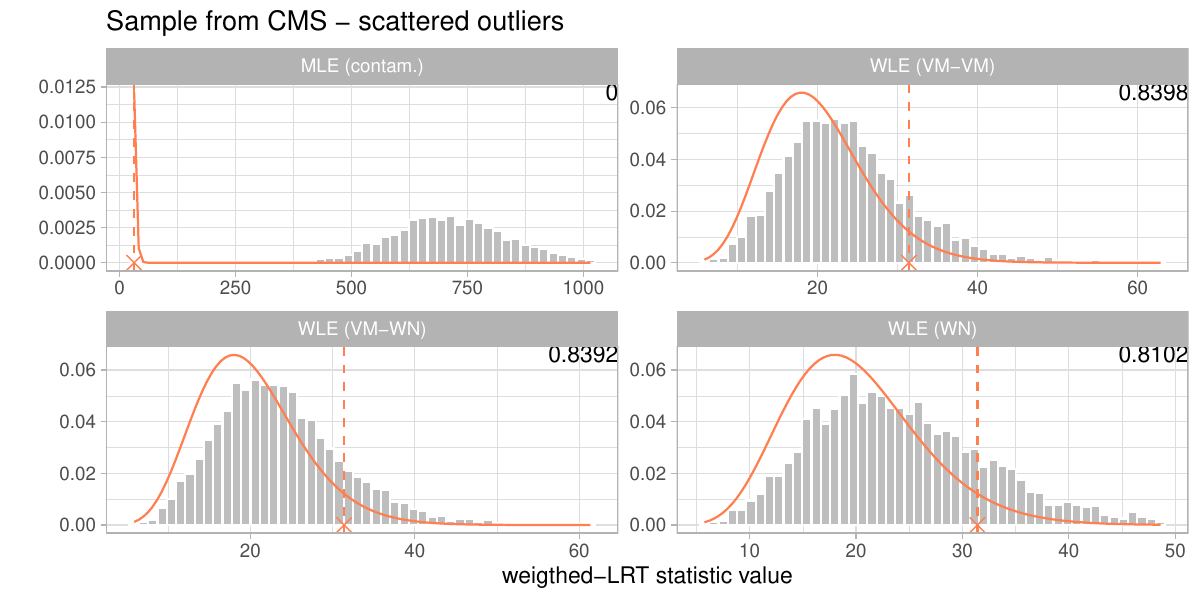}
	\caption{Monte Carlo simulations (five-dimensional). Sampling distributions for the WLRT (LRT for MLE under contamination) statistic and the $\chi^2-$ density superimosed. Empirical coverage at level 0.95 printed on the top-right of each subplot. Results are based on samples of size $n=300$ from a five-dimensional concentrated multivariate sine (CMS) distribution with scattered outliers.}
	\label{fig:wlrt-5d-cms-sc}
\end{figure}

\begin{table}[h]
	\begin{subtable}[h]{0.4\textwidth}
		\centering
		\begin{tabular}{l|l}
			\hline
			& coverage \\ \hline
			MLE (genuine) & 0.949   \\
			MLE (contam.) & 0        \\
			WLE (VM-VM)   & 0.6032   \\
			WLE (VM-WN)   & 0.6052   \\
			WLE (WN)      & 0.9584  
		\end{tabular}
		\caption{clustered outliers}
	\end{subtable}
	\hfill
	\begin{subtable}[h]{0.4\textwidth}
		\centering
		\begin{tabular}{l|l}
			\hline
			& coverage \\ \hline
			MLE (genuine) & 0.949   \\
			MLE (contam.) & 0        \\
			WLE (VM-VM)   & 0.8398   \\
			WLE (VM-WN)   & 0.8392   \\
			WLE (WN)      & 0.8102  
		\end{tabular}
		\caption{scattered outliers}
	\end{subtable}
	\caption{Monte Carlo simulations (five-dimensional). Empirical coverage for the LRT and WLRT for at level 0.95. Results are based on samples of size $n=300$ from a bivariate concentrated multivariate sine (CMS) distribution with contamination.}
	\label{tab:coverage-5d-cms}
\end{table}

% ---
% no contamination
% ---

\begin{table}[h]
	\begin{subtable}[h]{0.4\textwidth}
		\centering
		\begin{tabular}{lrrr}
		\hline
		& eff (VM-VM) & eff (VM-WN) & eff (WN) \\
		\hline
		$\hat{\vect{\mu}}$ & 0.9895 & 0.9895 & 1.0044 \\
		$\hat{\vect{\kappa}}$ & 0.2862 & 0.2861 & 0.9055 \\
		$\hat{\lambda}$ & 0.9042 & 0.9042 & 1.2014 \\
		\hline
		\end{tabular}
		\caption{bivariate case}
	\end{subtable}
	\hfill
	\begin{subtable}[h]{0.4\textwidth}
		\centering
		\begin{tabular}{lrrr}
		\hline
		& eff (VM-VM) & eff (VM-WN) & eff (WN) \\
		\hline
		$\hat{\vect{\mu}}$ & 0.9877 & 0.9865 & 0.9962 \\
		$\hat{\vect{\kappa}}$ & 0.4036 & 0.4010 & 0.9762 \\
		$\hat{\lambda}$ & 0.7798 & 0.7786 & 0.9375 \\
		\hline
		\end{tabular}
		\caption{five-dimensional case}
	\end{subtable}
	\caption{Monte Carlo simulations. Efficiency of the considered WLEs is evaluated taking the ratio between the Mean Squared Error of the MLE and of the WLE. Results are based on samples of size $n=300$ from a bivariate concentrated multivariate sine (CMS) distribution.}
	\label{tab:mse-cms-rt}
\end{table}

% Different parameters MLE approx MLE

\begin{figure}[h]
	\centering
	\includegraphics[width=0.9\textwidth]{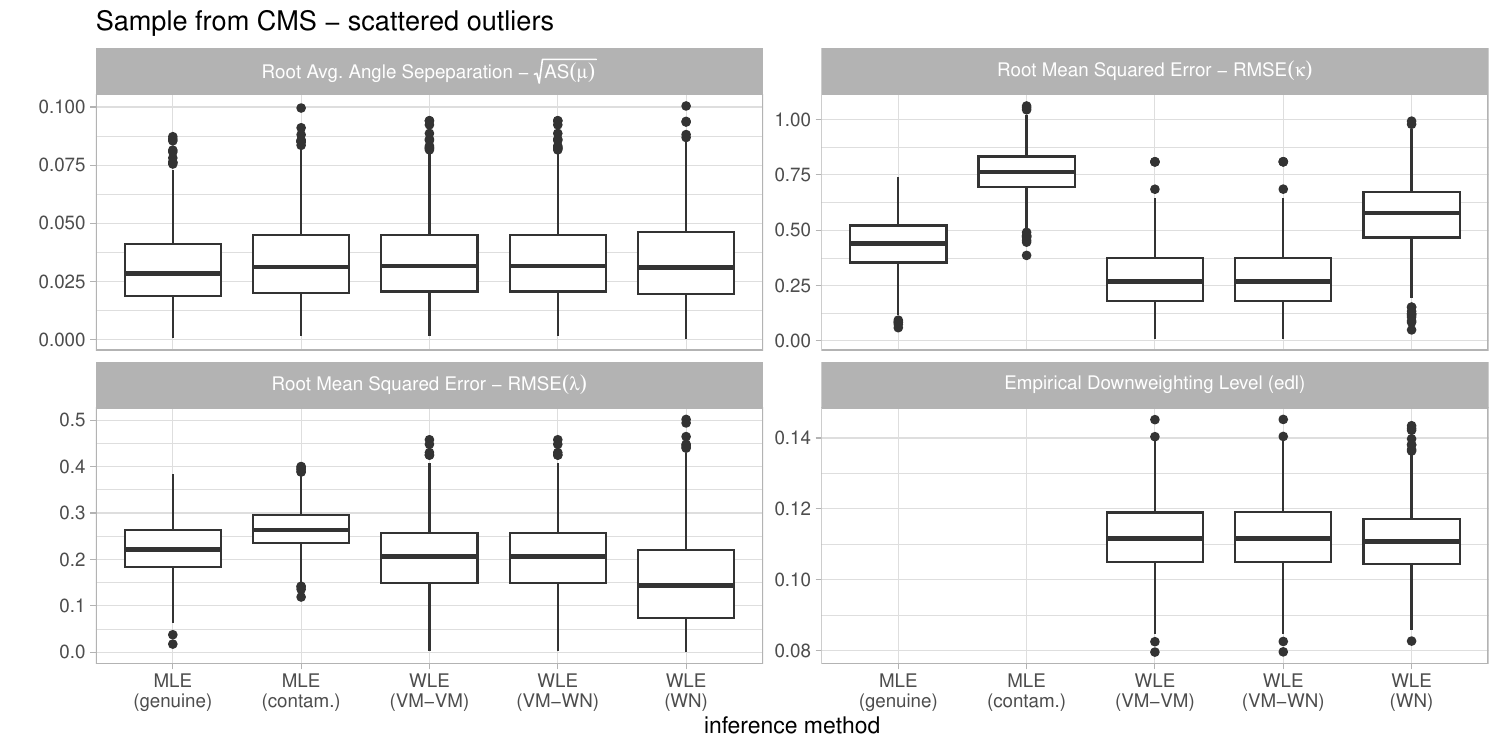}
	\caption{Monte Carlo simulation (bivariate). Empirical distributions of the estimation errors for parameters $\vect{\mu}$, $\vect{\kappa}$, and $\lambda$, together with the empirical downweighting level (edl). Different estimation methods are displayed along the $x-$axis. Results are based on samples of size $n=300$ from a bivariate CMS distribution with $\vect{\kappa}=(2, 2)$ and $\lambda=0.35$, with scattered outliers.}
	\label{fig:sim2d-bis-sc}
\end{figure}

\begin{figure}[h]
	\centering
	\includegraphics[width=0.9\textwidth]{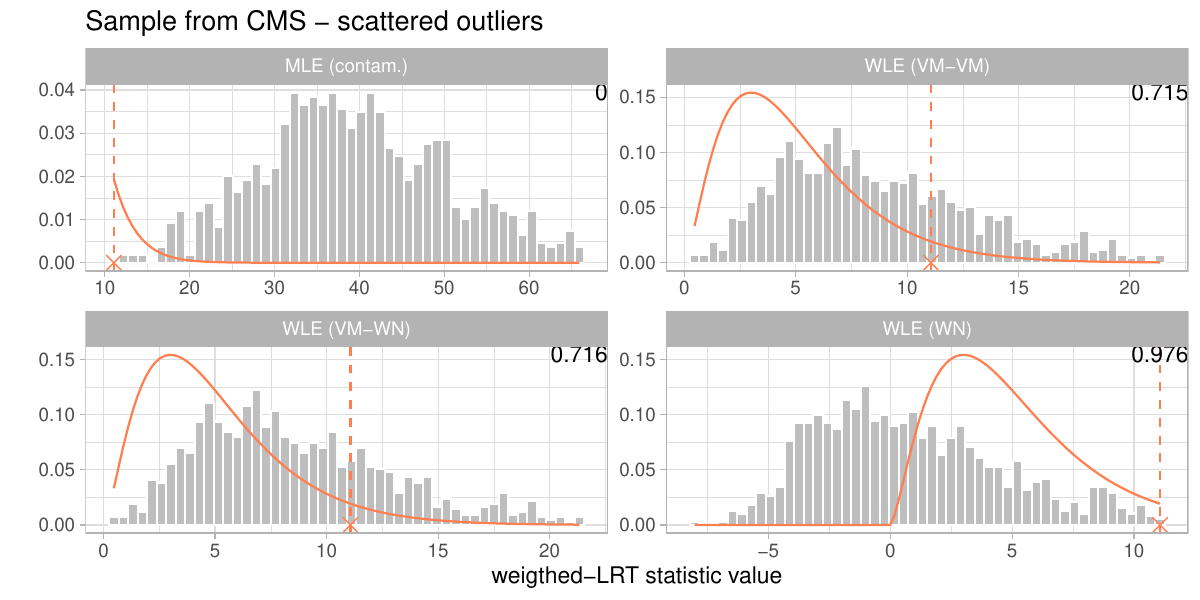}
	\caption{Monte Carlo simulations (bivariate). Sampling distributions for the WLRT (LRT for MLE under contamination) statistic and the $\chi^2-$ density superimosed. Empirical coverage at level 0.95 printed on the top-right of each subplot. Results are based on samples of size $n=300$ from a bivariate CMS distribution with $\vect{\kappa}=(2, 2)$ and $\lambda=0.35$, with scattered outliers.}
	\label{fig:wlrt-bis-sc}
\end{figure}

\begin{figure}[h]
	\centering
	\includegraphics[width=0.9\textwidth]{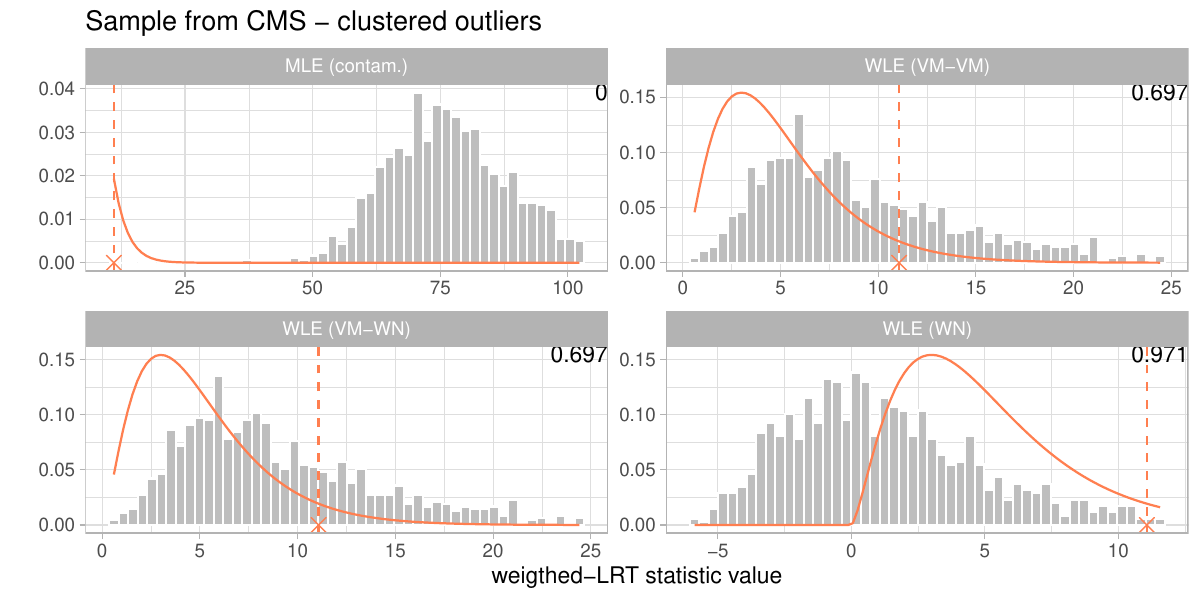}
	\caption{Monte Carlo simulations (bivariate). Sampling distributions for the WLRT (LRT for MLE under contamination) statistic and the $\chi^2-$ density superimosed. Empirical coverage at level 0.95 printed on the top-right of each subplot. Results are based on samples of size $n=300$ from a bivariate CMS distribution with $\vect{\kappa}=(2, 2)$ and $\lambda=0.35$, with scattered outliers.}
	\label{fig:wlrt-bis-gr}
\end{figure}

\clearpage

\bibliographystyle{plainnat}
\bibliography{biblio}

\end{document}